\documentclass{aa}  

\usepackage{soul}
\usepackage{ulem}
\usepackage{cancel}
\usepackage{graphicx}	
\usepackage{color}
\usepackage{natbib}
\usepackage{placeins}

\usepackage{txfonts}

\usepackage[breaklinks=true]{hyperref} 
\usepackage{algorithm}
\usepackage{amsfonts}
\usepackage{mathrsfs}
\usepackage{bm}

\usepackage{rotating}
\usepackage{color}
\usepackage{subfigure}
\usepackage{lineno}
\usepackage[toc,page]{appendix}
\usepackage[english]{babel}
\usepackage{xfrac}
\usepackage{comment}
\usepackage{tablefootnote}

\DeclareRobustCommand{\VAN}[3]{#2}
\let\VANthebibliography\thebibliography
\def\thebibliography{\DeclareRobustCommand{\VAN}[3]{##3}\VANthebibliography}

\usepackage{longtable}  
\usepackage{lscape}     
\usepackage{tabularx} 

\begin{document}

   \title{The deepest color-magnitude diagrams for the benchmark open cluster NGC\,2437 from Gaia and VVVX  
   }
   \titlerunning{The benchmark open cluster NGC2437} 

   \author{Tali Palma\inst{1}
          \and
          Mat\'ias G\'omez\inst{2}
          \and
          Dante Minniti\inst{2,3}
          \and
          Alan Montecinos\inst{2}
          \and
          Nicholas J. G. Cross\inst{4}
          \and
          Javier Alonso-Garc\'ia\inst{5}
          \and
          Daniel Majaess\inst{6}
          \and
          Bruno Dias\inst{2}
          \and 
          Maren Hempel\inst{2,7}
          \and
          Roberto K. Saito\inst{8}
          \and
          Joyce B. Pullen\inst{2}
          }

   \institute{Observatorio Astronómico, Universidad Nacional de Córdoba, Laprida 854, X5000BGR, Córdoba, Argentina\\
   \email{tpalma@unc.edu.ar}
   \and
   Instituto de Astrofísica, Depto. de Física y Astronomía, Facultad de Ciencias Exactas, Universidad Andrés Bello, Fernández Concha 700, Las Condes,
   Santiago, Chile
   \and
   Vatican Observatory, Vatican City State, V-00120, Italy
   \and
   Wide-Field Astronomy Unit, Institute for Astronomy, University of Edinburgh, Royal Observatory, Blackford Hill, Edinburgh EH9 3HJ, UK
   \and
   Centro de Astronomía (CITEVA), Universidad de Antofagasta, Av. Angamos 601, Antofagasta, Chile
   \and
   Mount Saint Vincent University, Halifax, Canada
   \and
   Max-Planck Institute for Astronomy, Heidelberg 69117, Koenigstuhl 17, Germany
   \and
   Departamento de Física, Universidade Federal de Santa Catarina, Trindade 88040-900, Florianópolis, SC, Brazil
   }

   \date{Received September 30, 20XX}

 \abstract
   {Deep photometry of Galactic star clusters provides one of the most powerful tools for determining their physical properties. In particular, in low Galactic latitude regions that suffer from heavy extinction and crowding, it is essential to have accurate optical and near-IR photometry for benchmark clusters.}
   {NGC\,2437 is the most extended star cluster in the near-IR footprint of the VISTA Variables in the Via Lactea Extended Survey (VVVX), covering more than one degree on the sky. In this paper, we aim to characterize the physical properties of NGC\,2437 using Gaia data release 3 (DR3) databases in the optical and the VVVX in the near-IR.}
   {We use Gaia DR3 proper motions (PMs) to select NGC\,2437 members in order to make optical and near-IR color-magnitude and color-color diagrams. We further exploited the newly constructed VVVX deep stack images to obtain the deepest near-IR color-magnitude diagram currently available for this cluster.}
   {Using the selected cluster members, we estimate the main physical parameters for NGC\,2437, including the mean parallax $\varpi = 0.608$ mas and PMs $(\mu_{\alpha*}, \mu_{\delta}) = (-3.85, 0.41)$ mas/yr. We also estimate a mean reddening of $E(J-K_s) = 0.059$ mag and an extinction of $A_{k_s} = 0.034$ mag for the NGC\,2437 field, with no significant differential reddening spread throughout the field.  A distance modulus of $(m-M) = 11.08$ mag is estimated, equivalent to a distance of $D=1644$ pc. This places NGC\,2437 at $z=115$ pc above the Galactic plane and at a galactocentric distance of $R_G = 9.24$ kpc. 
   We measure the structural parameters of the cluster by obtaining the core radius $r_c=10.79$ arcmin. We estimate the total absolute magnitude in the near-IR of $M_K = -4.91$ mag, equivalent to an optical absolute magnitude of $M_V = -3.70$ mag. 
   We measure a cluster mean age of $\rm{t} = 350$ Myr, using the PARSEC-COLIBRI isochrones for solar metallicity. 
   We also measure the binary fraction using the optical and near-IR color-magnitude diagrams. Based on Gaia DR3 data and VVVX deep stacks, we conclude that the NGC\,2437 binary fraction is 28.6\%.
   We also discuss the implications of the revised cluster parameters for the nearby open cluster NGC\,2425, the planetary nebula NGC\,2438, and the evolved OH/IR source OH\,231.8+04.2.}
   {The VVVX deep stacks increase the photometric depth by approximately $1.6$ mag in the $K_s$ band, nearly doubling the number of detected point sources and enabling significantly improved studies of stellar populations throughout the southern Galactic plane.}


   \maketitle

\section{Introduction}
Open clusters play an important role in Galactic research, and their study is one of the primary scientific goals of the VISTA Variables in the Via Lactea (VVV) Survey and its extension, the VVVX Survey \citep{Minniti2010,Saito2012,Saito2024}. These near-IR surveys have enabled a wide range of studies, including: 
the search for new open clusters at low Galactic latitudes \citep{Borissova2011,Borissova2014,Barba2015,Ivanov2017,Borissova2018,Borissova2020,Gupta2024}, 
the photometric and spectroscopic characterization of stellar members of young massive clusters \citep{Borissova2016,Borissova2019,Chene2012,Chene2013,Ramirez2014,Chene2015,Martins2019,Pena2022}, 
the study of variable stars in Galactic open clusters, including Cepheids, Miras, semiregular variables, and eclipsing binaries \citep{Majaess2011,Majaess2012,Dekany2015,Palma2016,Navarro2016,Medina2018,Molina2019,Medina2021,Majaess2024,Majaess2025}.

Color-magnitude diagrams (CMDs) in these previous studies have generally relied on single-epoch observations. In contrast, the present work makes use of newly generated deep stack images obtained by co-adding the highest-quality images acquired over the full temporal baseline of both surveys. This approach significantly increases the photometric depth and improves the signal-to-noise ratio, enabling the detection of substantially fainter stellar sources. As a first application of these new products, we analyze the deep stack corresponding to VVVX tile e0851, which contains interesting astronomical objects, namely the open clusters NGC\,2437 and NGC\,2425, the planetary nebula NGC\,2438, and the OH/IR source OH\,231.8+04.2. 

In this work, we perform an updated analysis of the benchmark open cluster NGC\,2437 (M\,46), which is one of the most extended open clusters located within the VVVX footprint. It appears as a prominent stellar overdensity in the new VVVX stellar density maps. This cluster has an apparent diameter close to one degree on the sky, much larger than the projected size of the Moon. 
The cluster lies in a relatively uncrowded region of the Galactic plane with relatively low interstellar extinction. 
Based on previous work, this is an intermediate-age nearby open cluster \citep[$\log{\rm{t}} = 8.4$, $D = 1.5$ kpc;][]{Sharma2006,Davidge2013,Jadhav2021}. 
It exhibits a well-defined binary sequence. The combination of Gaia DR3 astrometry with VVVX deep stack photometry allows us to construct very deep optical near-IR CMDs that extend toward the hydrogen-burning limit. Gaia proper motions (PMs) help secure and/or refine membership, while VVVX photometry enables the identification of significantly fainter objects. 

\begin{figure}
  \begin{center}
    \includegraphics[width=83mm]{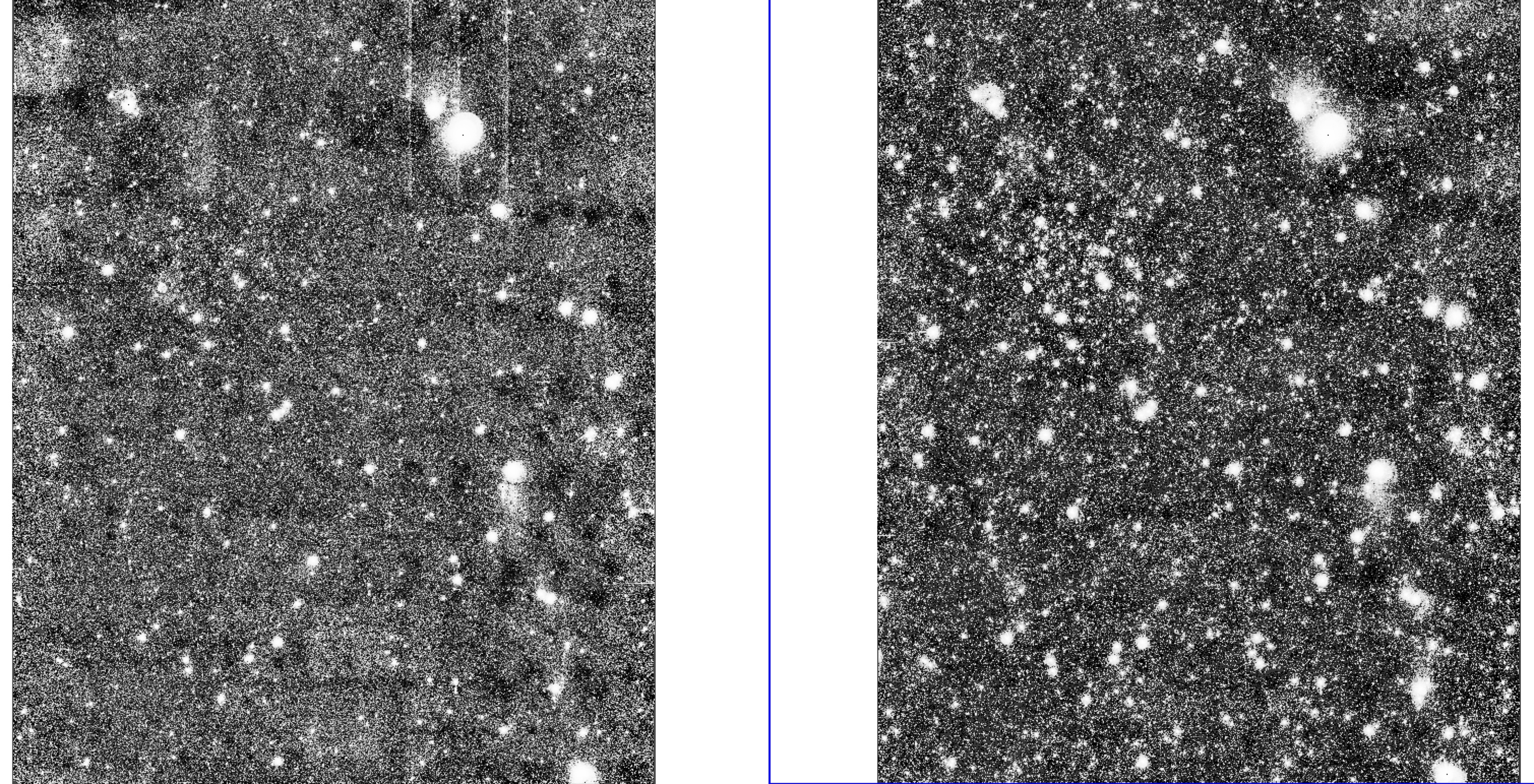}
    \caption{Deep stack images of the VVVX tile e0851 in the $J$ (left) and $K_s$ (right) passbands. The open cluster NGC\,2437 is located in the top left quadrant. }
    \label{Figure1}
  \end{center}
\end{figure}

\begin{figure}
  \begin{center}
    \includegraphics[width=82mm]{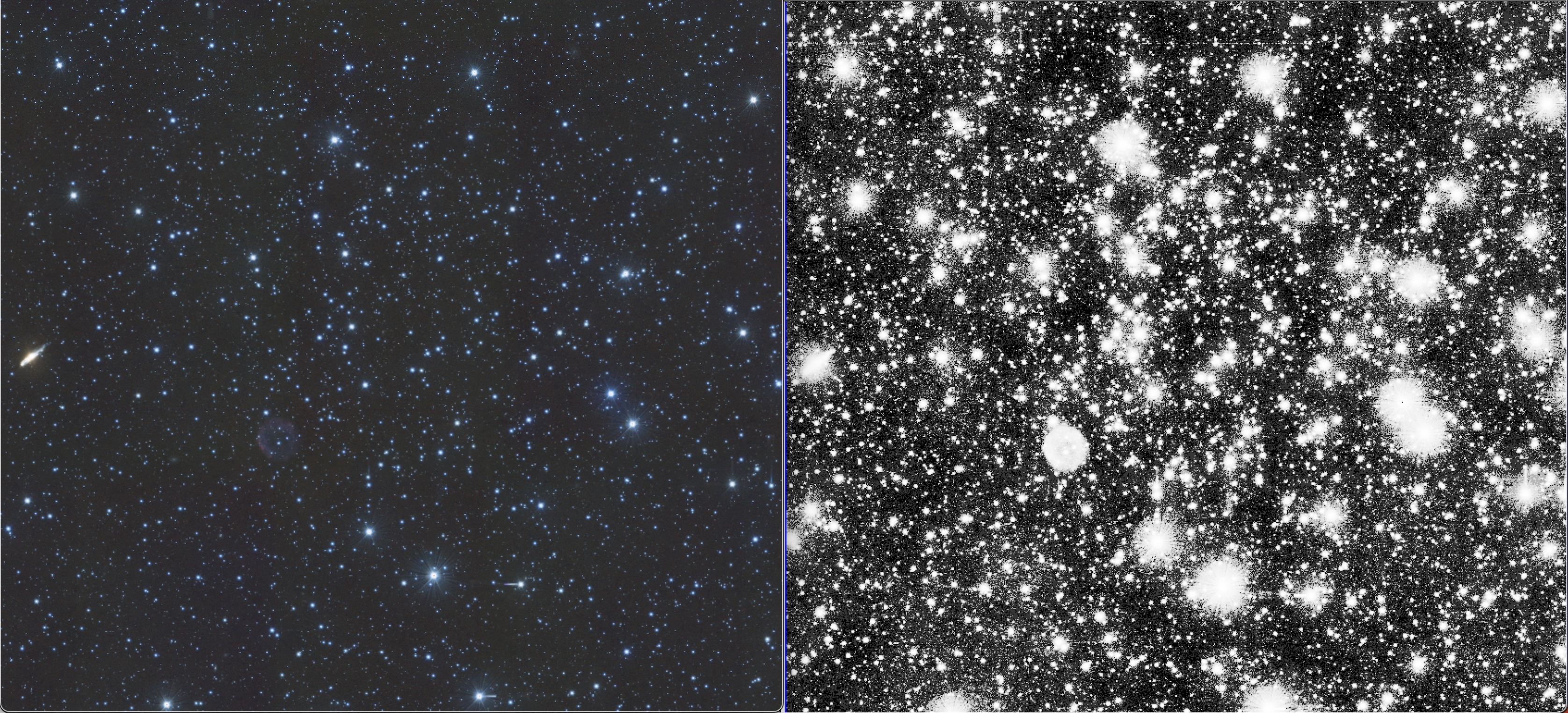}
    \caption{Open cluster NGC\,2437 in the near-IR. Zoom into the single epoch $JHK_s$ color image of NGC\,2437 (left) compared with the deep stack image in the $K_s$ passband (right). }
    \label{Figure2}
  \end{center}
\end{figure}

Fig. \ref{Figure1} shows the deep stacked $J$ and $K_s$ band images of the VVVX tile e0851, centered at $(RA,DEC)_{J2000}= $\,(07:39:45.16, -14:41:09.9), which according to \citet{Saito2024} contains two observations in the $J$ band, one observation in the $H$ band, and 30 observations in the $K_s$ band, spread between the years 2016 and 2025. After quality control was applied (see Sec. \ref{DeepStacks}), the deep stack input includes two sets of $J$ band and 22 sets of $K_s$ band. This explains the difference in the final quality of the deep stack images. NGC\,2437 is located off-center within tile e0851. Because this cluster extends beyond the VVVX footprint, only the central region ($r < 20$ arcmin) is completely covered by the survey and is used for the photometric analysis, whereas Gaia astrometry allows the identification of probable members out to projected radii of approximately $45$ arcmin. A comparison between a single epoch $JHK_s$ color image and a deep stacked image in the $K_s$ band of the same central region is shown in Fig. \ref{Figure2}.

Although extensively studied, NGC\,2437 remains an excellent benchmark open cluster due to its richness, proximity, and intermediate age. The availability of Gaia DR3 astrometry together with the recently released VVVX deep stacks provides an opportunity to revisit its structural and physical properties while extending the stellar census toward significantly deeper near-IR photometry than previously available. The main goals of this work are to derive an updated sample of cluster members, determine the structural and physical parameters of NGC\,2437, and assess the improvement provided by the new VVVX deep stack photometry compared to previous VVVX catalogs.

\section{Datasets and VVV near-infrared photometry}\label{data}

The VVV/VVVX surveys have systematically mapped the Galactic bulge and the southern Galactic plane in the near-IR using the VIRCAM instrument mounted on the VISTA 4m telescope at ESO Paranal Observatory between 2010 and 2023 \citep{Minniti2010,Saito2012}. 
During the course of the surveys, the photometric products have evolved from aperture photometry provided by the Cambridge Astronomical Survey Unit (CASU) to increasingly sophisticated point spread function (PSF) photometric catalogs, including ZKSPHOT \citep{Mauro2012}, 
DAOPHOT \citep{Contreras2017,Smith2018,Surot2019}, and most recently DoPhot \citep{Alonso2018,Alonso2026,Smith2025}. These successive improvements have substantially increased both the completeness and the photometric accuracy, particularly in crowded stellar fields. 
In this work, we used the latest DoPhot PSF photometric catalogs produced from  VVVX observations \citep{Alonso2026} together with the newly generated VVVX deep stack images. The latter are constructed by combining the highest-quality images available for each tile, resulting in substantially deeper photometry and improved photometric precision with respect to the single-epoch observations. 
The Gaia DR3 and VVVX catalogs were cross-matched using their equatorial coordinates, and only sources with reliable astrometric and photometric measurements were retained for the subsequent analysis presented in this work.

\section{The VVVX deep stacks}\label{DeepStacks}

With the completion of the VVVX survey, it is now possible to construct deep stacked images by combining the highest-quality observations obtained over the full temporal baseline of the survey.
In this paper we present the first scientific application of the VVVX deep stack products. These are being developed at the Wide Field Astronomy Unit (WFAU) of the Royal Observatory Edinburgh.
The stacked images are constructed using only the highest-quality exposures, rather than by co-adding all available observations.  Images affected by poor seeing, elongated PSFs, reflections, or high sky background were rejected prior to the stacking process. For tile e0851, the following quantitative quality-control criteria were applied prior to stacking: seeing thresholds of $1.0$\,arcsec and $2.0$\,arcsec ($K_s$ master and epoch tiles, respectively), $1.2$\,arcsec ($J$), and $1.1$\,arcsec ($H$); PSF ellipticity $< 0.2$ in all bands; and photometric zero-point within $0.15$ mag of the median value for $J$, $H$, and $K_s$ (master), and within $0.25$ mag for $K_s$ (epoch tiles). After applying these criteria, two $J$ band and 22 $K_s$ band observations were retained for the final deep stacks of tile e0851. The resulting stack images achieve a final PSF full width at half maximum (FWHM) of approximately $0.82$ arcsec in both $J$ and $K_s$ bands, with PSF ellipticities of $0.087$ and $0.05$, respectively, representing a significant improvement not only in depth, but also in angular resolution with respect to individual single-epoch observations.

Here we assess the gain in photometric depth provided by the deep stacks using VVVX tile e0851 as a representative example. 
Fig. \ref{FigureA1} illustrates the improvement achieved with the deep stacks. This figure shows the number of point sources detected using
DoPhot PSF photometry of a single chip for VVVX tile e0851, located in the Galactic plane ($l$ = 231.510 deg, $b$ = 3.701 deg). 
This figure compares the PSF photometry of single images in filters $J$ and $K_s$ \citep[black histogram,][]{Alonso2026}, with the deep stacks of a single $K_s$ band filter (light gray histogram). 

The deep stack photometry reaches approximately 1.6 mag deeper in the $K_s$ band than the single-epoch catalogs, increasing the number of detected point sources by a factor of two.
This results in the detection of 121423 more point sources.
This number is representative of the photometric yields for tiles located in the Galactic plane. 
In addition, the  photometry from the deep stack exhibits smaller photometric uncertainties. 
The improvement is expected to be even greater in low-density fields, where crowding is less severe, whereas the gain is likely to be smaller in the innermost regions of the Galactic bulge, where source confusion rather than photon noise dominates the detection limits. In addition to extending the limiting magnitude, the deep stack products improve the completeness and photometric precision in crowded stellar fields, allowing significantly cleaner CMDs at faint magnitudes.

\subsection{Lucky imaging}

The final depth achievable by deep stacks depends primarily on the quality of the individual input images.
The faintest sources are preferentially detected in the images obtained under the best seeing conditions.
 
The Lucky imaging technique has been successfully used to identify faint sources in astronomical images at different wavelengths \citep[see][]{Law2006,Lodieu2009}. 
This concept has not yet been fully exploited within the VVVX database, but the approach may prove fruitful, especially at low latitude fields where the photometric completeness is limited by crowding.
This strategy is particularly advantageous in the crowded inner regions of the Galaxy, where improvements in both angular resolution and photometric depth translate into significantly higher photometric completeness.
The image-selection strategy adopted for the deep stacks is conceptually similar to the lucky imaging approach in the sense that only the highest-quality exposures are combined.

\subsection{Rotation differential imaging}

The VISTA telescope at ESO Cerro Paranal Observatory is an altazimuth telescope, so the images at different epochs are taken with different instrument rotations. 
This means that the image diffraction pattern that is seen in the bright stars in the individual image is rotated arbitrarily from image to image of the same tile. 
This rotation averages out the diffraction pattern in the co-added deep stack images, 
with the bright stars resulting in point sources exhibiting extended smooth halos. 
Source catalogs extracted from the individual images required additional cleaning because of spurious detections produced by diffraction spikes around bright stars. The photometry of the individual images had to be cleaned up as a result of the detection of false sources due to the diffraction pattern of bright stars. 
For example, when two bright stars are close together, some of their diffraction features intersect, resulting in the detection of false point sources. 
This is not necessarily the case in the deep co-added images, where the diffraction pattern has been averaged out (as long as many images are combined). 
In a sense, this effect is similar to the rotation differential imaging (RDI) technique applied to detect planetary companions around bright stars \citep[see][]{Soummer2011,Soummer2012,Sanghi2022}. An additional advantage of combining images acquired at different instrument position angles is the suppression of diffraction spikes around bright stars.

\subsection{Roman}

The deep VVVX near-IR stacks are also important in the context of the future Galactic Plane Survey to be carried out by the Nancy Grace Roman Space Telescope at similar wavelengths (e.g., Paladini et al. 2022). 
The high-resolution images acquired with the Wide Field Instrument (WFI) at the Roman Space Telescope would reach faint magnitudes in short exposure times (e.g., $J > 23$ mag, $K_s > 21$ mag in integrations of approximately one minute). 
Unfortunately, stars brighter than $K_s < 15$ mag would be saturated even in the shortest exposures. 
The deep stacks extend the photometry of our survey to fainter magnitudes, providing a wider dynamical range of overlap (4-5 mag) with the photometry of the Roman Space Telescope, thereby providing a substantial common magnitude range for photometric calibration.
This overlap is important not only for photometric calibration, but also because it provides a temporal baseline of nearly two decades for future PM studies of sources with $15<K_s<19.5$ mag across the Galactic plane and bulge.

\section{The NGC\,2437 color-magnitude and color-color diagrams}

\begin{figure}
 \begin{center}
    \includegraphics[width=62mm]{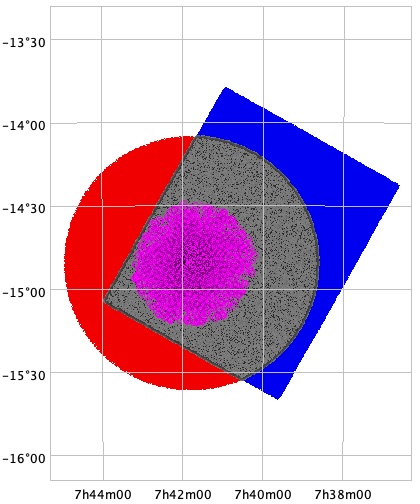}
    \caption{Map of the coverage of the VVVX tile e0851 (blue rectangle) compared with the sample of NGC\,2437 members selected using the Gaia DR3 proper motions (pink circle). The red and gray circle has a diameter of 45’ for reference.  }
    \label{Figure3}
 \end{center}
\end{figure}

The spatial distribution of the Gaia–VVVX field and matched sources in the vicinity of NGC\,2437, together with the footprint of the VVVX tile e0851, is shown in Fig. \ref{Figure3}. The figure illustrates the location of the cluster within the tile and highlights that only the central region is fully covered by the VVVX observations, while the outermost members extend beyond the survey footprint.

Reliable membership determination is essential in low Galactic latitude fields, where the stellar density is dominated by foreground and background Galactic populations. Therefore, we used Gaia DR3 PMs and parallaxes to isolate probable members of NGC\,2437 from the surrounding field population. An initial selection was performed in the PM plane, followed by a parallax criterion centered on the mean cluster parallax. The resulting sample was finally inspected in the Gaia CMD to remove obvious contaminants while preserving the cluster sequence.

\begin{figure}
 \begin{center}
    \includegraphics[width=55mm]
    {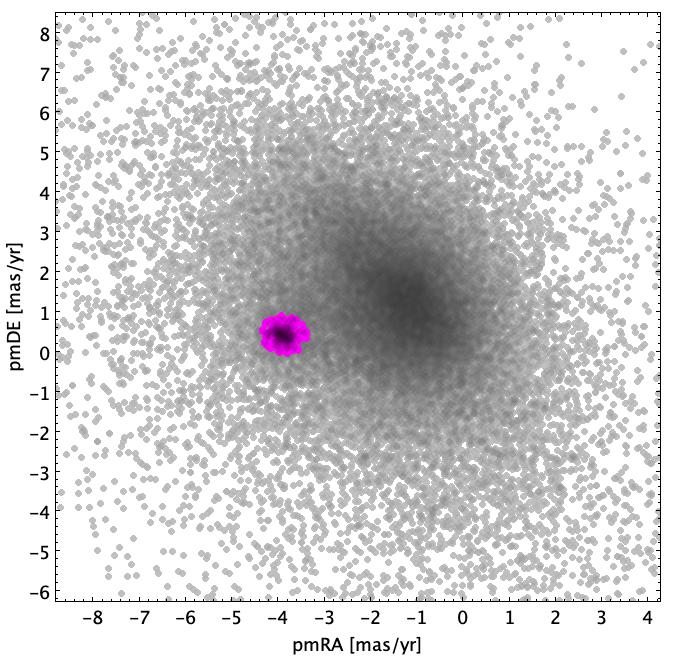}
    \caption{Proper motion selection of cluster members within $r<20$ arcmin from the center of NGC\,2437.  }
    \label{Figure4}
 \end{center}
\end{figure}

Fig. \ref{Figure4} illustrates the adopted PM selection for stars within $r<20$ arcmin of the cluster center. The relatively compact distribution in the proper-motion plane demonstrates the effectiveness of the astrometric criteria in isolating the cluster population from the surrounding field stars. The PM selection was performed within a circle with radius of $0.5$\,mas/yr centered on the mean cluster motion. Subsequently, a parallax criterion of $\varpi = 0.6 \pm 0.1$\,mas was applied to further reject field contaminants.

\begin{figure*}
 \begin{center}
    \includegraphics[width=180mm]{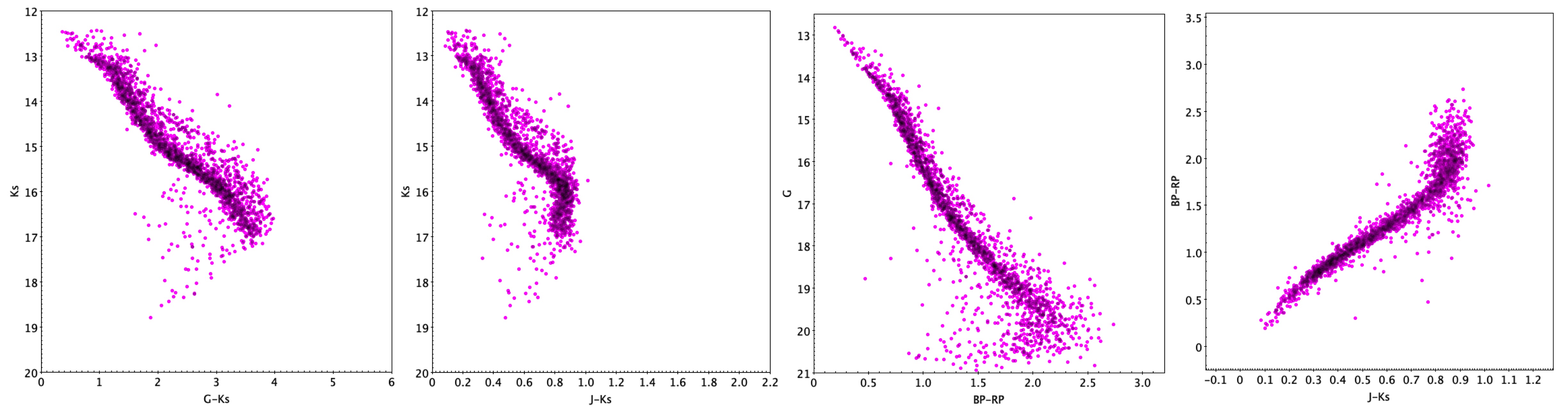}
    \caption{Near-IR color-magnitude and color-color diagrams for the Gaia proper motion selected sources within $r<20'$ from the center of NGC\,2437.  }
    \label{Figure5}
 \end{center}
\end{figure*}

The resulting optical and near-IR CMDs, together with the color–color diagram, are presented in Fig. \ref{Figure5}. The selected members define a narrow and well-populated main sequence, confirming the high quality of the astrometric membership selection.  The tight main sequence of the cluster is consistent with the relatively low foreground extinction and the modest level of differential reddening across the cluster field, as shown in the color-color diagram (right panel of Fig. \ref{Figure5}). These clean photometric sequences provide the basis for the determination of the cluster structural parameters and the isochrone fitting presented in the following sections. \\

Although Gaia DR3 astrometry provides a robust membership determination over a wide magnitude range, PMs and parallaxes become increasingly incomplete toward the faintest magnitudes. To investigate the low-luminosity stellar population revealed by the VVVX deep stacks, we therefore complement the astrometric selection with a statistical field-star decontamination procedure in order to discriminate the background and foreground star fields from the
observed cluster CMDs. Following \citet{Palma2016,Palma2019,Minniti2017,Minniti2018}, a nearby comparison field was selected approximately one degree from the cluster center, covering a similar projected area, and chosen to have comparable Galactic latitude, stellar density, and foreground reddening. 

\begin{figure*}
 \begin{center}
    \includegraphics[width=160mm]{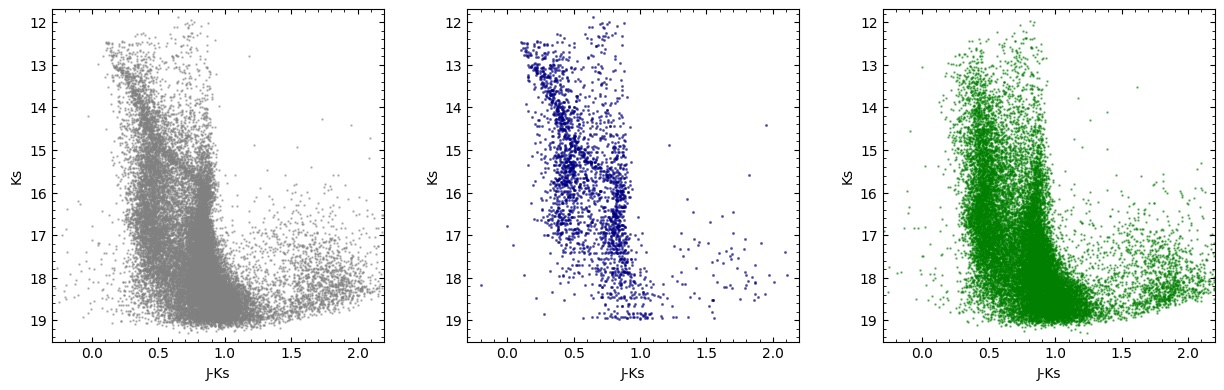}
    \caption{Near-IR color-magnitude diagram for the deep stacks for the whole VVVX tile e0851 (left), compared with the sources within $r<20$ arcmin from the center of NGC\,2437 (middle) and with a background field located about 1 degree away (right), separated using statistical decontamination techniques (see text). Note that these CMDs are much deeper that those shown in Figure 5, that are based only on Gaia proper motion selected sources. }
    \label{Figure6}
 \end{center}
\end{figure*}

Fig. \ref{Figure6} shows the $K_s$ $\rm{vs}$ $J-K_s$ CMD for the deep stacks of the entire VVVX tile e0851 (left panel), the central region of NGC\,2437 ($r<20'$, middle panel), and the comparison field located approximately one degree away from the cluster (right panel). 
Due to the use of deep stack photometry, these diagrams reach significantly fainter magnitudes than the Gaia-selected CMDs presented in Fig. \ref{Figure5}. The comparison field exhibits typical stellar populations expected along this line of sight in the Galactic plane \citep[see Fig. 6 in][]{Alonso2026}. Three prominent sequences can be distinguished:
the distant reddened main-sequence dwarfs (with mean $J-K_s \sim 0.45$ mag), 
the unreddened nearby M-dwarfs (with mean $J-K_s \sim 0.85$ mag),
and the population dominated by background galaxies (with $J-K_s > 1.2$ mag and a wide color spread).

The comparison between the cluster and control field confirms that the prominent cluster main sequence is clearly distinguishable from the underlying Galactic population. Statistical decontamination therefore provides an effective complement to the Gaia-based membership selection at magnitudes where astrometric information becomes incomplete.

\section{Main physical parameters of NGC\,2437}\label{Parameters}

Using Gaia DR3 astrometry together with VVVX deep stack photometry, we derive updated structural and physical parameters for NGC\,2437. The cluster center was estimated at $RA =$ 07:41:49.08 and $DEC =$ -14:49:46.54. The mean cluster parallax and PMs are $\varpi = 0.608 \pm 0.041$ mas, $\mu_{\alpha*} = -3.85 \pm 0.13$ mas/yr and $\mu_{\delta} = 0.41 \pm 0.11$ mas/yr.  The measured parallax corresponds to a heliocentric distance of $D = 1.64 \pm 0.11$ kpc (without applying the Bayesian correction of \citet{Bailer2021} ), equivalent to a distance modulus of  $(m-M) = 11.08$ mag. These values are supported by the subset of Gaia DR3 members with reliable parallaxes and radial velocities, which are predominantly bright cluster members. This sample yields a mean radial velocity of $RV = +46.3 \pm 5.6$ km/s. The derived astrometric parameters are in good agreement with recent determinations based on Gaia DR3 \citep[e.g.,][]{Hunt2023,Hunt2024}. The structural parameters of the cluster were derived by fitting a King model \citep{king1962} to the projected radial density profile of the Gaia-selected members, which yields a core radius of $r_c= 10.79  \pm 0.66$ arcmin (corresponding to $r_c = 5.15 \pm 0.47$ pc).

The mean reddening and extinction derived for the cluster are $E(J-K_s) = 0.059$ mag and $A_{k_s} = 0.034$ mag. For a consistency check, we computed the mean reddening $E(BP-RP)$ from the Gaia DR3 photometry of the cluster members. Following the extinction law of \citet{WangChen2019}, the above adopted values are in excellent agreement with the value derived from the Gaia photometry.
We adopted a metallicity of $\rm{[Fe/H]} = -0.08 \pm 0.16$ and solar abundance ratios from the high-resolution spectroscopic analysis of \citet{Ray2022}. 
The cluster age was estimated using the PARSEC-COLIBRI  isochrones \citep{Bressan2012,Marigo2017}. We derived an age of $\log{\rm{t}} = 8.55$, equivalent to $\rm{t} = 355$ Myr.  For the adopted age and metallicity, the PARSEC-COLIBRI isochrones predict a main-sequence turn-off mass of $\rm M_{TO} \sim 3.05\, M_{\odot}$, corresponding to an early A-type star. The derived structural and physical parameters are summarized in Table \ref{tab:params}.

\section{The NGC\,2437 binary fraction}\label{Binaries}

We used the Gaia DR3 + VVVX matched catalog to estimate the binary fraction of NGC\,2437. 
The binary population is most clearly identified in the $(G-K_s, K_s)$ CMD, because the broad optical - near-IR color baseline provides a better separation between single-star and unresolved binary sequences. The resulting CMD clearly reveals the NGC\,2437 cluster MS, as well as a parallel binary MS approximately $0.7$ mag above the single-star MS, as expected for unresolved equal-mass binaries. This well-defined binary sequence enables a direct estimate of the cluster binary fraction. 
The cluster MS is detected down to $K_s > 17.0$ mag, $G > 20.5$ mag.  We conclude that residual field contamination is small, based on the low number of sources located blueward and below the cluster MS.

We counted the total number of detected sources ($\rm N_{all}$), the stars located along the cluster MS ($\rm N_{MS}$), the candidate unresolved binaries located above the MS ($\rm N_{ bin}$), and the likely field contaminants below the MS ($\rm N_{ bkg}$), following a procedure similar to that described by \citet{Jadhav2021}. We obtained $\rm N_{all} = 3319$, $\rm N_{MS} = 2293$ (69\%), $\rm N_{bin} = 655$ (20\%), and $\rm N_{bkg} = 319$ (10\%). Following the observational approach adopted in this work, we define the photometric binary fraction as $f_{\rm bin} = \rm N_{bin}/N_{MS}$, where these are the numbers of stars assigned to the binary and single-star MS loci, respectively. We obtain a binary fraction of $0.28$ for the cluster. 

Note that this photometric method is sensitive only to unresolved binaries with mass ratios $q \sim 1$. Binaries with lower mass ratios ($q\ll 1$) produce smaller photometric offsets and are not counted as part of $\rm N_{bin}$. Similarly, wide binaries that are spatially resolved at the distance of NGC\,2437 are not included. The quoted binary fraction therefore represents a lower limit to the total binary fraction of the cluster.

The analysis covers a specific range of magnitudes, covering spectral types from early A (near the turn-off) to late K at the faint end, mostly GK stars. Although the astrometric quality of Gaia DR3 is degraded significantly for $G > 18$ mag, which may introduce some contamination in the faint end of our sample, its effect on the overall binary fraction estimate is expected to be small given the low number of sources in this regime.

We also inspected the Gaia DR3 renormalized unit weight error (RUWE) distribution of the candidate binary population. A dozen candidates exhibit high values (RUWE $> 1.4$), as expected for systems whose unresolved orbital motion or multiplicity may affect the Gaia astrometric solution. However, the RUWE distribution does not provide sufficiently clear evidence to establish a distinct population of wide binaries or higher-order multiples in the present sample.

Adopting a more restrictive selection, taking into account only sources within the tight sequence seen in the color-color diagrams, excluding stars with larger photometric uncertainties or possible blends, the number of main sequence stars drops to $\rm N_{cc} = 2150$, but the inferred binary fraction remains essentially unchanged because the binaries drop accordingly. 
Despite extending the photometry by more than two magnitudes relative to previous studies, we obtain a binary fraction consistent with the value reported by \citet{Jadhav2021}. This agreement indicates that the inferred binary fraction is robust against the improved photometric depth provided by the VVVX deep stacks.

\section{The open cluster NGC2425}\label{OCNGC2425}

\begin{figure}
  \begin{center}
    \includegraphics[width=90mm]{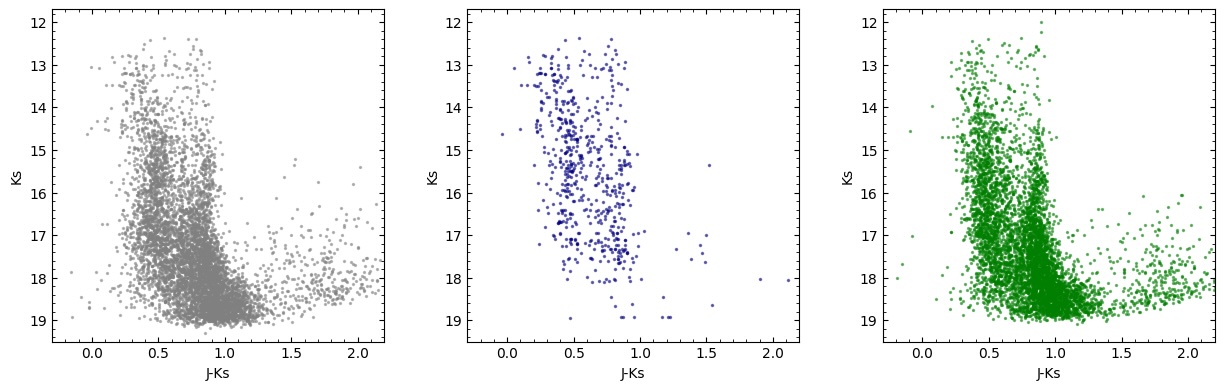}
    \caption{Near-IR CMD for NGC2425 using the deep stacks of VVVX tile e0851 (left panel), compared with a nearby background field (right panel). The resulting statistically decontaminated CMD is shown in the middle panel. The tight cluster main sequence can be seen as an excess of blue stars in the left panel. }
         \label{Figure7}
  \end{center}
\end{figure}

The VVVX deep stacks also allow us to investigate the more distant open cluster NGC\,2425, located within the same tile as NGC\,2437.
NGC\,2425 is an intermediate-age open cluster located approximately $0.5^{\circ}$ from NGC\,2437 within the same VVVX tile, but at a substantially larger heliocentric distance.  Its central coordinates are $(RA, DEC)_{J2000} = $\, (07:36:00.9, -14:46:14), and its Gaia DR3 parallax of $0.257 \pm 0.005$ mas corresponds to a heliocentric distance of approximately $4$ kpc.
Fig. \ref{Figure7} shows the statistically decontaminated near-IR CMD for NGC\,2425 using the deep stacks compared to a background field located half a degree away from the cluster.
The excess of blue stars due to the tight cluster MS can be clearly observed on the left side of this CMD at the brightest magnitudes ($K_s < 17$ mag).

Although Gaia DR3 astrometry was also explored for NGC\,2425, at its substantially larger distance, the Gaia sample becomes incomplete at significantly brighter magnitudes than VVVX deep stack photometry, making statistical decontamination a more effective approach for this cluster.
This example illustrates the capabilities of the VVVX  deep stacks. While they extend the stellar census of nearby clusters such as NGC\,2437 toward lower masses, their increased photometric depth is expected to be even more beneficial for the study of more distant Galactic open clusters.
        
\section{The planetary nebula NGC\,2438}\label{PNNGC2438}

\begin{figure}
  \begin{center}
    \includegraphics[width=90mm]{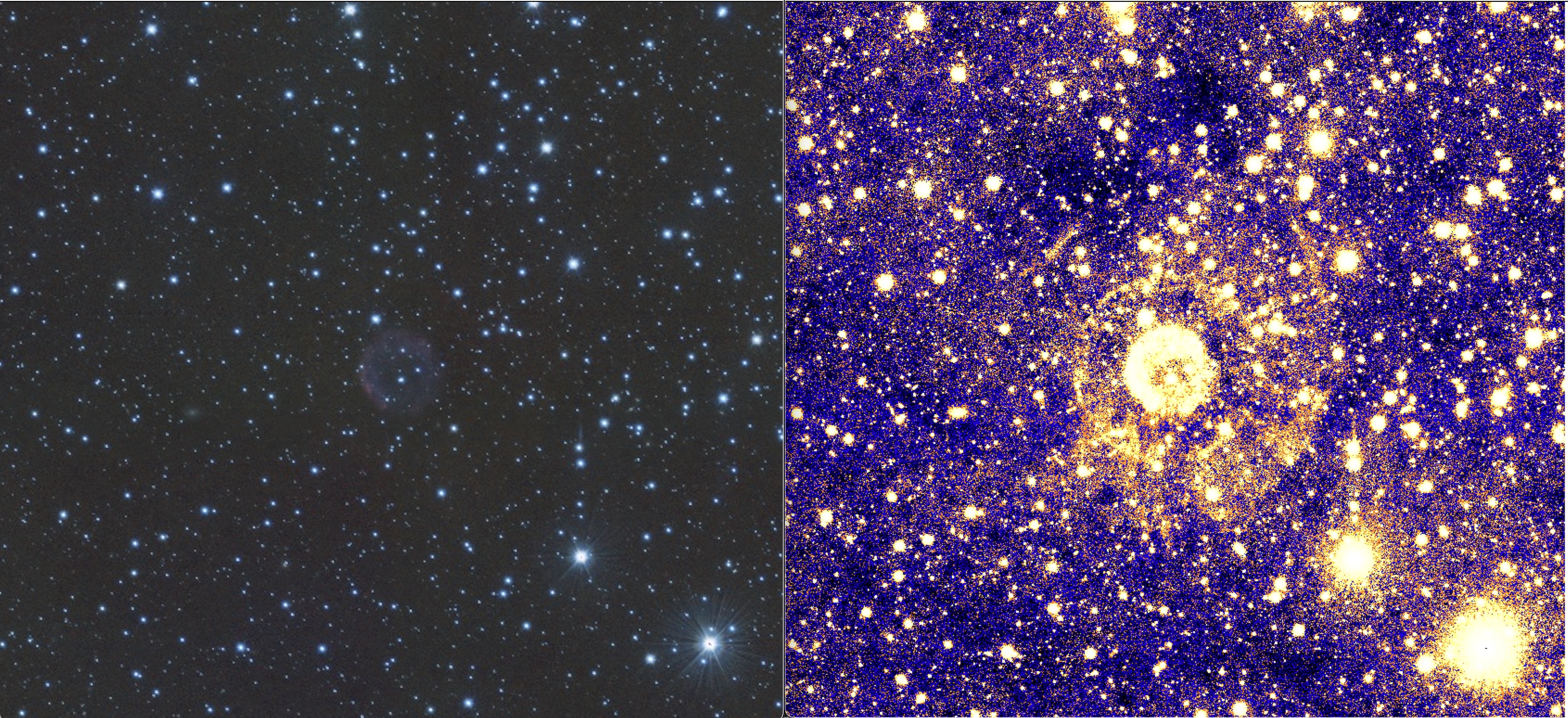}
    \caption{Planetary nebula NGC\,2438 in the near-IR. Zoom into the single epoch $JHK_s$ color image (left), compared with the deep stack $J$ band image of this planetary nebula. Note the multiple nested shell structure of this object seen in the $J$ band.}
    \label{Figure8}
  \end{center}
\end{figure}

The planetary nebula NGC\,2438 (PN G231.8+04.1) is another interesting object located within VVVX tile e0851.
This object exhibits a multiple-shell structure in the deep optical images taken from the ground and space \citep[e.g.,][]{Corradi2000,Ottl2014}. 
Fig. \ref{Figure8} shows the zoomed VVVX $J$ and $K_s$ band images of this planetary nebula (PN), with a total diameter of 11’ centered at $(RA,DEC)_{J2000}$ =\, (07:41: 50.52, -14:44:07.5), $l = 231.510$ deg, $b = 3.701$ deg.
This figure reveals the same multiple-shell morphology in the near-IR.

The central star for this PN has $G = 17.08$ mag, $BP-RP = -0.05$ mag, and $A_V = 0.14$ mag \citep[from the Gaia EDR3 data analysis of][]{Gonzalez2021}. 
We measured the near-IR PSF magnitudes for this central star: 
$J = 16.75  \pm 0.01$ mag, 
$H = 16.39  \pm  0.02$ mag, 
$K_s = 16.29  \pm  0.05$ mag using the VVVX images.
This PN is particularly interesting because it is located within the field of the open cluster NGC\,2437 \citep{Majaess2007}.

According to \citet{Kiss2008}, the radial velocity of the PN, $RV = 78 \pm 2$ km/s, rules out a physical association with NGC\,2437, whose mean radial velocity is $48.5$ km/s.
Indeed, the PN velocity is more than $5 \sigma$ away from our mean radial velocity value of  $RV = +46.3 \pm 5.6$ km/s, strongly arguing against cluster membership. 
The Gaia database for the central star of this PN yields $\mu_{\alpha*} = -5.625$ mas\,yr$^{-1}$, $\mu_{\delta} = 1.519$ mas\,yr$^{-1}$, and $\varpi = 1.38  \pm   0.22$ mas, corresponding to an inverse-parallax distance of approximately $720$ pc. 
\citet{Ottl2014} derive a total extinction $E(B-V) = 0.16$ mag, estimating a distance of $D = 1.9 \pm  0.2$ kpc, again suggesting nonmembership of the open cluster NGC\,2437 (although the distance constraint is less compelling than the radial-velocity evidence because of the larger distance uncertainties). We therefore conclude that NGC\,2438 is a chance line-of-sight projection toward NGC\,2437 rather than a physical cluster member.

\section{The OH/IR source OH\,231.8+04.2}\label{OH/IR}

The OH/IR source OH\,231.8+04.2, also known as QX Pup and associated with the Calabash Nebula, is a bright evolved binary system located within VVVX tile e0851
\citep{Sanchez2004}. 
This remarkable object has been studied in a variety of astrophysical contexts, including as a water-vapor and silicon monoxide maser source \citep[e.g.,][]{Kim2019}, a protoplanetary nebula with jets and shocks \citep{Bujarrabal2002, Meakin2003}, a preplanetary nebula with a precessing jet \citep[e.g.,][]{Sahai2005}, a post-AGB star with a compact circumstellar disk \citep[e.g.,][]{Matsuura2006,Sanchez2022}, 
an oxygen-rich binary Mira with $P = 554$ d undergoing mass transfer \citep[e.g.,][]{Kastner1998,Sanchez2004,Sanchez2018}, and the driving source of a bipolar nebular outflow \citep{Reipurth1987,Morris1987,Hekkert1988,Shure1995}, which motivated complex modeling \citep[e.g.,][]{Balick2017,Bowers1984,Balick2017}.

This source lies well within the extension of the open cluster NGC\,2437, about $10’.67$ ($640".3$) from its center, and its radial velocity agrees with that of the cluster, providing strong evidence for a physical association \citep{Bowers1984,Sanchez2004}. 
Establishing a cluster membership of OH\,231.8+04.2 is important because
the precise distance influences its determination of age and the mass of the central binary components. 
Given that we present updated estimates of the relevant cluster parameters, such as size, age, and distance, it is important to confirm the physical association of the OH/IR source OH\,231.8+04.2 with NGC\,2437. 
Our updated cluster parameters are fully consistent with recent determinations, which have been used to compute the OH\,231.8+04.2 parameters, and no significant revision is necessary for this object. 
Specifically, \citet{Sanchez2022} estimated that the mass of the QX Pup progenitor is $\rm M \sim 3.5 M_{\odot}$ with a companion A\,05 of initial mass $\rm M \sim  2 M_{\odot}$. 
Their assumed distance $D = 1.3$ kpc and age $\rm{t} = 225-250$ Myr adopted for these calculations are broadly consistent within the uncertainties.
However, our larger distance of $D = 1.64$ kpc would change the energy budget previously estimated within the nebula. 
The larger distance implies luminosities approximately $0.45$ mag brighter than previously assumed, which may require a modest upward revision of the inferred stellar masses.
In addition, our distance estimate is also consistent with the value adopted by the hydrodynamical models of \citet{Balick2017}.
The estimated progenitor mass of QX Pup ($\sim 3.5\,\rm M_{\odot}$) is slightly larger than the cluster turn-off mass of $\sim 3.05\,\rm M_{\odot}$, derived from our isochrone fitting. This small difference ($\sim 15\%$) is physically consistent with cluster membership: a star with an initial mass slightly above the current turn-off would have already evolved off the main sequence and entered the AGB or post-AGB phase on a timescale compatible with the cluster age. Given the uncertainties in both age determination and stellar mass estimates from evolutionary models, the progenitor mass of OH 231.8+04.2 is in good agreement with the cluster turn-off, providing further support for the physical association of this system with NGC\,2437.

\begin{figure}
  \begin{center}
    \includegraphics[width=90mm]{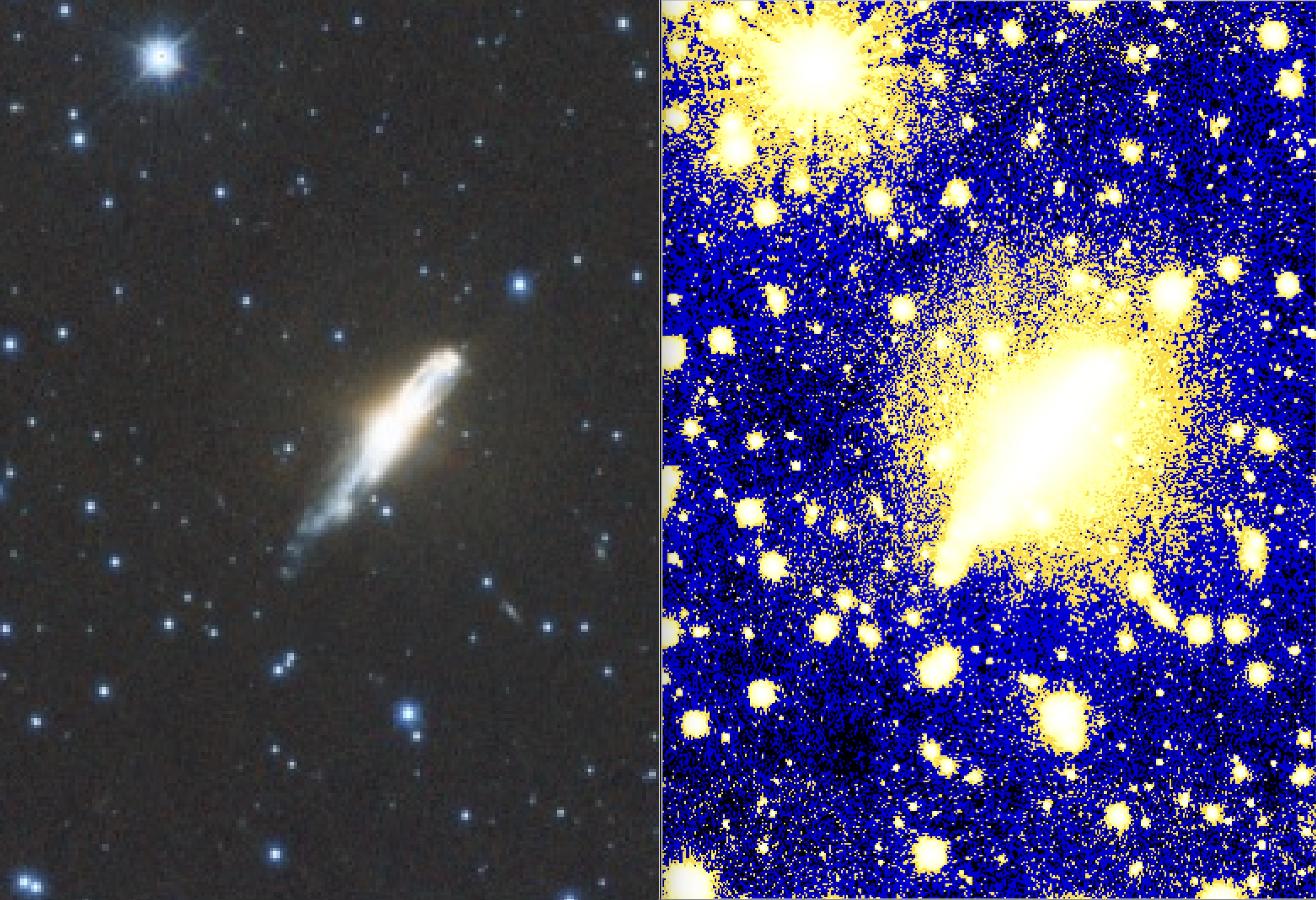}
    \caption{The OH/IR source OH\,231.8+04.2 in the near-IR. Zoom into the deep stack images of the OH/IR source OH\,231.8+04.2 in the single epoch $JHK_s$ color image (left), compared with the deep stack in the $K_s$ passband (right). }
    \label{Figure9}
  \end{center}
\end{figure}

 \begin{table}
\centering
\caption{Fundamental parameters of NGC\,2437.}
\label{tab:params}
\begin{tabular}{ll}
\hline
Parameter & Value \\
\hline
$RA$  & 07:41:49.08 \\
$DEC$  & $-14$:49:46.54 \\
$\mu_{\alpha*}$ [mas/yr] & $-3.85 \pm 0.13 $ \\
$\mu_{delta}$ [mas/yr] & $0.41 \pm 0.11$ \\
$\varpi$ [mas] & $0.608 \pm 0.041$  \\
$A_{K_s}$ [mag] & $0.034 \pm 0.003$ \\
$A_G$ [mag] & $0.230 \pm 0.006 $ \\
$(m-M)_0$ [mag] & $11.08 \pm 0.15 $ \\
$D_{\odot}$ [kpc] & $1.64 \pm 0.11$ \\
Age [Myr] &  $355 \pm 90$ \\
$R_{\rm GC}$ [kpc] & $9.24 \pm 0.05$ \\
$Z_{\rm Gal}$ [kpc] & $0.115 \pm 0.005 $ \\
$RV$ [km/s] & $46.3 \pm 5.6$ \\
$r_c$ [pc] & $5.15 \pm 0.47$ \\
$E(J-K_s)$ [mag] & $0.059 \pm 0.003$ \\
$M_{K_s}$ [mag] & $-4.91 \pm  0.09$ \\
$M_V$ [mag] & $-3.70 \pm  0.09 $ \\
\hline
\end{tabular}
\end{table}
\section{Conclusions}\label{conclusions}

We presented the first scientific application of the VVVX deep stack project, demonstrating the gain in photometric depth and source completeness achieved by combining the highest-quality VVVX observations. As a showcase of these new data products, we analyzed the benchmark open cluster NGC\,2437.
Using deep stack photometry together with Gaia DR3, we derived updated estimates of reddening, distance, age, luminosity, structural parameters, and mean proper motion of the benchmark open cluster NGC\,2437 (Table\,\ref{tab:params}). 

The VVVX deep stacks extend the limiting magnitude by approximately $1.6$ mag in the $K_s$ band, approximately doubling the number of detected point sources in typical Galactic-plane fields, while simultaneously improving the photometric precision and completeness at faint magnitudes.
The photometric binary fraction is estimated to be approximately 30\%, consistent with previous determinations despite extending the CMD by more than two magnitudes toward fainter stars.

Finally, improved photometry allowed us to investigate several astrophysically interesting objects within the same VVVX tile, including the distant open cluster NGC\,2425, the planetary nebula NGC\,2438, and the evolved OH/IR source OH\,231.8+04.2. The new cluster parameters provide additional support for the nonmembership of NGC\,2438 and remain consistent with the interpretation of OH\,231.8+04.2 as a member of NGC\,2437.
The deep stack products will provide an important legacy dataset for future Galactic surveys, particularly in combination with Gaia and the Nancy Grace Roman Space Telescope, enabling improved studies of stellar populations and proper motions throughout the southern Galactic plane and bulge.

\begin{acknowledgements}\\

{We gratefully acknowledge the use of data from the ESO Public Survey program IDs 179.B-2002 and 198.B-2004 taken with the VISTA telescope and data products from the Cambridge Astronomical Survey Unit. \\
T.P. gratefully acknowledge support from Secretaría de Ciencia y Tecnología de la Universidad Nacional de Córdoba (SECYT-UNC, Res.
258/53). \\
M.G. and D.M. gratefully acknowledge support  from Fondecyt Project No. 1220724.\\
D.M. also gratefully acknowledges support from the Center for Astrophysics and Associated Technologies CATA by the ANID BASAL projects ACE210002 and FB210003.
}
\end{acknowledgements}

\bibliographystyle{aa} 
\bibliography{Biblio.bib} 

@ARTICLE{Alonso2018,
       author = {{Alonso-Garc{\'\i}a}, Javier and {Saito}, Roberto K. and {Hempel}, Maren and {Minniti}, Dante and {Pullen}, Joyce and {Catelan}, M{\'a}rcio and {Ramos}, Rodrigo Contreras and {Cross}, Nicholas J.~G. and {Gonzalez}, Oscar A. and {Lucas}, Philip W. and {Palma}, Tali and {Valenti}, Elena and {Zoccali}, Manuela},
        title = "{Milky Way demographics with the VVV survey. IV. PSF photometry from almost one billion stars in the Galactic bulge and adjacent southern disk}",
      journal = {\aap},
         year = 2018,
        month = oct,
       volume = {619},
          eid = {A4},
        pages = {A4},
          doi = {10.1051/0004-6361/201833432},
archivePrefix = {arXiv},
       eprint = {1808.06139},
 primaryClass = {astro-ph.GA},
       adsurl = {https://ui.adsabs.harvard.edu/abs/2018A&A...619A...4A}
}

@ARTICLE{Alonso2026,
       author = {{Alonso-Garc{\'\i}a}, Javier and {Hempel}, Maren and {Saito}, Roberto K. and {Minniti}, Dante and {Cross}, Nicholas J.~G. and {Anais}, Jorge and {Borissova}, Jura and {Catelan}, M{\'a}rcio and {Fern{\'a}ndez-Trincado}, Jos{\'e} G. and {Garro}, Elisa R. and {Guo}, Zhen and {Lucas}, Philip W. and {Navarro}, Mar{\'\i}a G. and {Obasi}, Casmir O. and {Smith}, Leigh C.},
        title = "{A near-infrared stellar atlas of the Galactic plane from the VVVX survey}",
      journal = {\aap},
         year = 2026,
        month = feb,
       volume = {706},
          eid = {A301},
        pages = {A301},
          doi = {10.1051/0004-6361/202557979},
archivePrefix = {arXiv},
       eprint = {2601.19708},
 primaryClass = {astro-ph.GA},
       adsurl = {https://ui.adsabs.harvard.edu/abs/2026A&A...706A.301A}
}

@ARTICLE{Bailer2021,
       author = {{Bailer-Jones}, C.~A.~L. and {Rybizki}, J. and {Fouesneau}, M. and {Demleitner}, M. and {Andrae}, R.},
        title = "{Estimating Distances from Parallaxes. V. Geometric and Photogeometric Distances to 1.47 Billion Stars in Gaia Early Data Release 3}",
      journal = {\aj},
         year = 2021,
        month = mar,
       volume = {161},
       number = {3},
          eid = {147},
        pages = {147},
          doi = {10.3847/1538-3881/abd806},
archivePrefix = {arXiv},
       eprint = {2012.05220},
 primaryClass = {astro-ph.SR},
       adsurl = {https://ui.adsabs.harvard.edu/abs/2021AJ....161..147B}
}

@ARTICLE{Balick2017,
       author = {{Balick}, Bruce and {Frank}, Adam and {Liu}, Baowei and {Huarte-Espinosa}, Mart{\'\i}n},
        title = "{Models of the Hydrodynamic Histories of Post-AGB Stars. I. Multiflow Shaping of OH 231.8+04.2}",
      journal = {\apj},
         year = 2017,
        month = jul,
       volume = {843},
       number = {2},
          eid = {108},
        pages = {108},
          doi = {10.3847/1538-4357/aa77f0},
archivePrefix = {arXiv},
       eprint = {1706.01516},
 primaryClass = {astro-ph.SR},
       adsurl = {https://ui.adsabs.harvard.edu/abs/2017ApJ...843..108B}
}

@ARTICLE{Barba2015,
       author = {{Barb{\'a}}, R.~H. and {Roman-Lopes}, A. and {Nilo Castell{\'o}n}, J.~L. and {Firpo}, V. and {Minniti}, D. and {Lucas}, P. and {Emerson}, J.~P. and {Hempel}, M. and {Soto}, M. and {Saito}, R.~K.},
        title = "{Hundreds of new cluster candidates in the VISTA Variables in the V{\'\i}a L{\'a}ctea survey DR1}",
      journal = {\aap},
         year = 2015,
        month = sep,
       volume = {581},
          eid = {A120},
        pages = {A120},
          doi = {10.1051/0004-6361/201424048},
archivePrefix = {arXiv},
       eprint = {1505.02764},
 primaryClass = {astro-ph.SR},
       adsurl = {https://ui.adsabs.harvard.edu/abs/2015A&A...581A.120B}
}

@ARTICLE{Borissova2020,
       author = {{Borissova}, J. and {Kurtev}, R. and {Amarinho}, N. and {Alonso-Garc{\'\i}a}, J. and {Ram{\'\i}rez Alegr{\'\i}a}, S. and {Bernal}, S. and {Medina}, N. and {Chen{\'e}}, A.-N. and {Ivanov}, V.~D. and {Lucas}, P.~W. and {Minniti}, D.},
        title = "{Small-scale star formation as revealed by VVVX galactic cluster candidates}",
      journal = {\mnras},
         year = 2020,
        month = dec,
       volume = {499},
       number = {3},
        pages = {3522-3533},
          doi = {10.1093/mnras/staa3045},
archivePrefix = {arXiv},
       eprint = {2009.14372},
 primaryClass = {astro-ph.SR},
       adsurl = {https://ui.adsabs.harvard.edu/abs/2020MNRAS.499.3522B}
}

@ARTICLE{Borissova2018,
       author = {{Borissova}, J. and {Ivanov}, V.~D. and {Lucas}, P.~W. and {Kurtev}, R. and {Alonso-Garcia}, J. and {Ram{\'\i}rez Alegr{\'\i}a}, S. and {Minniti}, D. and {Froebrich}, D. and {Hempel}, M. and {Medina}, N. and {Chen{\'e}}, A.-N. and {Kuhn}, M.~A.},
        title = "{New Galactic star clusters discovered in the disc area of the VVVX survey}",
      journal = {\mnras},
         year = 2018,
        month = dec,
       volume = {481},
       number = {3},
        pages = {3902-3920},
          doi = {10.1093/mnras/sty2354},
archivePrefix = {arXiv},
       eprint = {1811.01944},
 primaryClass = {astro-ph.GA},
       adsurl = {https://ui.adsabs.harvard.edu/abs/2018MNRAS.481.3902B}
}

@ARTICLE{Borissova2014,
       author = {{Borissova}, J. and {Chen{\'e}}, A.-N. and {Ram{\'\i}rez Alegr{\'\i}a}, S. and {Sharma}, S. and {Clarke}, J.~R.~A. and {Kurtev}, R. and {Negueruela}, I. and {Marco}, A. and {Amigo}, P. and {Minniti}, D. and {Bica}, E. and {Bonatto}, C. and {Catelan}, M. and {Fierro}, C. and {Geisler}, D. and {Gromadzki}, M. and {Hempel}, M. and {Hanson}, M.~M. and {Ivanov}, V.~D. and {Lucas}, P. and {Majaess}, D. and {Moni Bidin}, C. and {Popescu}, B. and {Saito}, R.~K.},
        title = "{New galactic star clusters discovered in the VVV survey. Candidates projected on the inner disk and bulge}",
      journal = {\aap},
         year = 2014,
        month = sep,
       volume = {569},
          eid = {A24},
        pages = {A24},
          doi = {10.1051/0004-6361/201322483},
archivePrefix = {arXiv},
       eprint = {1406.7051},
 primaryClass = {astro-ph.GA},
       adsurl = {https://ui.adsabs.harvard.edu/abs/2014A&A...569A..24B}
}

@ARTICLE{Borissova2011,
       author = {{Borissova}, J. and {Bonatto}, C. and {Kurtev}, R. and {Clarke}, J.~R.~A. and {Pe{\~n}aloza}, F. and {Sale}, S.~E. and {Minniti}, D. and {Alonso-Garc{\'\i}a}, J. and {Artigau}, E. and {Barb{\'a}}, R. and {Bica}, E. and {Baume}, G.~L. and {Catelan}, M. and {Chen{\`e}}, A.~N. and {Dias}, B. and {Folkes}, S.~L. and {Froebrich}, D. and {Geisler}, D. and {de Grijs}, R. and {Hanson}, M.~M. and {Hempel}, M. and {Ivanov}, V.~D. and {Kumar}, M.~S.~N. and {Lucas}, P. and {Mauro}, F. and {Moni Bidin}, C. and {Rejkuba}, M. and {Saito}, R.~K. and {Tamura}, M. and {Toledo}, I.},
        title = "{New Galactic star clusters discovered in the VVV survey}",
      journal = {\aap},
         year = 2011,
        month = aug,
       volume = {532},
          eid = {A131},
        pages = {A131},
          doi = {10.1051/0004-6361/201116662},
archivePrefix = {arXiv},
       eprint = {1106.3045},
 primaryClass = {astro-ph.GA},
       adsurl = {https://ui.adsabs.harvard.edu/abs/2011A&A...532A.131B}
}

@ARTICLE{Borissova2019,
       author = {{Borissova}, Jura and {Roman-Lopes}, Alexandre and {Covey}, Kevin and {Medina}, Nicolas and {Kurtev}, Radostin and {Roman-Zuniga}, Carlos and {Kuhn}, M.~A. and {Contreras Pe{\~n}a}, Carlos and {Lucas}, Philip and {Ramirez Alegria}, Sebastian and {Minniti}, Dante and {Kounkel}, Marina and {Stringfellow}, Guy and {Barb{\'a}}, Rodolfo H. and {Su{\'a}rez}, Genaro},
        title = "{The G305 Star-forming Region. I. Newly Classified Hot Stars}",
      journal = {\aj},
         year = 2019,
        month = jul,
       volume = {158},
       number = {1},
          eid = {46},
        pages = {46},
          doi = {10.3847/1538-3881/ab276b},
       adsurl = {https://ui.adsabs.harvard.edu/abs/2019AJ....158...46B}
}

@ARTICLE{Borissova2016,
       author = {{Borissova}, J. and {Ram{\'\i}rez Alegr{\'\i}a}, S. and {Alonso}, J. and {Lucas}, P.~W. and {Kurtev}, R. and {Medina}, N. and {Navarro}, C. and {Kuhn}, M. and {Gromadzki}, M. and {Retamales}, G. and {Fernandez}, M.~A. and {Agurto-Gangas}, C. and {Chen{\'e}}, A.-N. and {Minniti}, D. and {Contreras Pena}, C. and {Catelan}, M. and {Decany}, I. and {Thompson}, M.~A. and {Morales}, E.~F.~E. and {Amigo}, P.},
        title = "{Young Stellar Clusters Containing Massive Young Stellar Objects in the VVV Survey}",
      journal = {\aj},
         year = 2016,
        month = sep,
       volume = {152},
       number = {3},
          eid = {74},
        pages = {74},
          doi = {10.3847/0004-6256/152/3/74},
archivePrefix = {arXiv},
       eprint = {1610.08474},
 primaryClass = {astro-ph.GA},
       adsurl = {https://ui.adsabs.harvard.edu/abs/2016AJ....152...74B}
}

@ARTICLE{Bowers1984,
       author = {{Bowers}, P.~F. and {Morris}, M.},
        title = "{The three-dimensional stucture of a circumstellar maser.}",
      journal = {\apj},
         year = 1984,
        month = jan,
       volume = {276},
        pages = {646-652},
          doi = {10.1086/161652},
       adsurl = {https://ui.adsabs.harvard.edu/abs/1984ApJ...276..646B}
}

@ARTICLE{Bressan2012,
       author = {{Bressan}, Alessandro and {Marigo}, Paola and {Girardi}, L{\'e}o. and {Salasnich}, Bernardo and {Dal Cero}, Claudia and {Rubele}, Stefano and {Nanni}, Ambra},
        title = "{PARSEC: stellar tracks and isochrones with the PAdova and TRieste Stellar Evolution Code}",
      journal = {\mnras},
         year = 2012,
        month = nov,
       volume = {427},
       number = {1},
        pages = {127-145},
          doi = {10.1111/j.1365-2966.2012.21948.x},
archivePrefix = {arXiv},
       eprint = {1208.4498},
 primaryClass = {astro-ph.SR},
       adsurl = {https://ui.adsabs.harvard.edu/abs/2012MNRAS.427..127B}
}

@ARTICLE{Bujarrabal2002,
       author = {{Bujarrabal}, V. and {Alcolea}, J. and {S{\'a}nchez Contreras}, C. and {Sahai}, R.},
        title = "{HST observations of the protoplanetary nebula OH 231.8+4.2: The structure of the jets and shocks}",
      journal = {\aap},
         year = 2002,
        month = jul,
       volume = {389},
        pages = {271-285},
          doi = {10.1051/0004-6361:20020455},
       adsurl = {https://ui.adsabs.harvard.edu/abs/2002A&A...389..271B}
}

@ARTICLE{Chene2012,
       author = {{Chen{\'e}}, A.-N. and {Borissova}, J. and {Clarke}, J.~R.~A. and {Bonatto}, C. and {Majaess}, D.~J. and {Moni Bidin}, C. and {Sale}, S.~E. and {Mauro}, F. and {Kurtev}, R. and {Baume}, G. and {Feinstein}, C. and {Ivanov}, V.~D. and {Geisler}, D. and {Catelan}, M. and {Minniti}, D. and {Lucas}, P. and {de Grijs}, R. and {Kumar}, M.~S.~N.},
        title = "{Massive open star clusters using the VVV survey. I. Presentation of the data and description of the approach}",
      journal = {\aap},
         year = 2012,
        month = sep,
       volume = {545},
          eid = {A54},
        pages = {A54},
          doi = {10.1051/0004-6361/201219064},
archivePrefix = {arXiv},
       eprint = {1206.6104},
 primaryClass = {astro-ph.SR},
       adsurl = {https://ui.adsabs.harvard.edu/abs/2012A&A...545A..54C}
}

@ARTICLE{Chene2013,
       author = {{Chen{\'e}}, A.-N. and {Borissova}, J. and {Bonatto}, C. and {Majaess}, D.~J. and {Baume}, G. and {Clarke}, J.~R.~A. and {Kurtev}, R. and {Schnurr}, O. and {Bouret}, J.-C. and {Catelan}, M. and {Emerson}, J.~P. and {Feinstein}, C. and {Geisler}, D. and {de Grijs}, R. and {Herv{\'e}}, A. and {Ivanov}, V.~D. and {Kumar}, M.~S.~N. and {Lucas}, P. and {Mahy}, L. and {Martins}, F. and {Mauro}, F. and {Minniti}, D. and {Moni Bidin}, C.},
        title = "{Massive open star clusters using the VVV survey. II. Discovery of six clusters with Wolf-Rayet stars}",
      journal = {\aap},
         year = 2013,
        month = jan,
       volume = {549},
          eid = {A98},
        pages = {A98},
          doi = {10.1051/0004-6361/201220107},
archivePrefix = {arXiv},
       eprint = {1211.2801},
 primaryClass = {astro-ph.SR},
       adsurl = {https://ui.adsabs.harvard.edu/abs/2013A&A...549A..98C}
}

@ARTICLE{Chene2015,
       author = {{Chen{\'e}}, A.-N. and {Ram{\'\i}rez Alegr{\'\i}a}, S. and {Borissova}, J. and {O'Leary}, E. and {Martins}, F. and {Herv{\'e}}, A. and {Kuhn}, M. and {Kurtev}, R. and {Consuelo Amigo Fuentes}, P. and {Bonatto}, C. and {Minniti}, D.},
        title = "{Massive open star clusters using the VVV survey. IV. WR 62-2, a new very massive star in the core of the VVV CL041 cluster}",
      journal = {\aap},
         year = 2015,
        month = dec,
       volume = {584},
          eid = {A31},
        pages = {A31},
          doi = {10.1051/0004-6361/201525958},
archivePrefix = {arXiv},
       eprint = {1510.02539},
 primaryClass = {astro-ph.SR},
       adsurl = {https://ui.adsabs.harvard.edu/abs/2015A&A...584A..31C}
}

@ARTICLE{Contreras2017,
       author = {{Contreras Ramos}, R. and {Zoccali}, M. and {Rojas}, F. and {Rojas-Arriagada}, A. and {G{\'a}rate}, M. and {Huijse}, P. and {Gran}, F. and {Soto}, M. and {Valcarce}, A.~A.~R. and {Est{\'e}vez}, P.~A. and {Minniti}, D.},
        title = "{Proper motions in the VVV Survey: Results for more than 15 million stars across NGC 6544}",
      journal = {\aap},
         year = 2017,
        month = dec,
       volume = {608},
          eid = {A140},
        pages = {A140},
          doi = {10.1051/0004-6361/201731462},
archivePrefix = {arXiv},
       eprint = {1709.07919},
 primaryClass = {astro-ph.GA},
       adsurl = {https://ui.adsabs.harvard.edu/abs/2017A&A...608A.140C}
}

@ARTICLE{Corradi2000,
       author = {{Corradi}, R.~L.~M. and {Sch{\"o}nberner}, D. and {Steffen}, M. and {Perinotto}, M.},
        title = "{A hydrodynamical study of multiple-shell planetaries . I. NGC 2438}",
      journal = {\aap},
         year = 2000,
        month = feb,
       volume = {354},
        pages = {1071-1085},
       adsurl = {https://ui.adsabs.harvard.edu/abs/2000A&A...354.1071C}
}

@ARTICLE{Davidge2013,
       author = {{Davidge}, T.~J.},
        title = "{The Open Cluster NGC 2437 (Messier 46)}",
      journal = {\pasp},
         year = 2013,
        month = feb,
       volume = {125},
       number = {924},
        pages = {115},
          doi = {10.1086/669823},
archivePrefix = {arXiv},
       eprint = {1301.6806},
 primaryClass = {astro-ph.GA},
       adsurl = {https://ui.adsabs.harvard.edu/abs/2013PASP..125..115D}
}

@ARTICLE{Dekany2015,
       author = {{D{\'e}k{\'a}ny}, I. and {Minniti}, D. and {Hajdu}, G. and {Alonso-Garc{\'\i}a}, J. and {Hempel}, M. and {Palma}, T. and {Catelan}, M. and {Gieren}, W. and {Majaess}, D.},
        title = "{Discovery of a Pair of Classical Cepheids in an Invisible Cluster Beyond the Galactic Bulge}",
      journal = {\apjl},
         year = 2015,
        month = jan,
       volume = {799},
       number = {1},
          eid = {L11},
        pages = {L11},
          doi = {10.1088/2041-8205/799/1/L11},
archivePrefix = {arXiv},
       eprint = {1412.8658},
 primaryClass = {astro-ph.GA},
       adsurl = {https://ui.adsabs.harvard.edu/abs/2015ApJ...799L..11D}
}

@ARTICLE{Gonzalez2021,
       author = {{Gonz{\'a}lez-Santamar{\'\i}a}, I. and {Manteiga}, M. and {Manchado}, A. and {Ulla}, A. and {Dafonte}, C. and {L{\'o}pez Varela}, P.},
        title = "{Planetary nebulae in Gaia EDR3: Central star identification, properties, and binarity}",
      journal = {\aap},
         year = 2021,
        month = dec,
       volume = {656},
          eid = {A51},
        pages = {A51},
          doi = {10.1051/0004-6361/202141916},
archivePrefix = {arXiv},
       eprint = {2109.12114},
 primaryClass = {astro-ph.GA},
       adsurl = {https://ui.adsabs.harvard.edu/abs/2021A&A...656A..51G}
}

@ARTICLE{Gupta2024,
       author = {{Gupta}, Akash and {Ivanov}, Valentin D. and {Preibisch}, Thomas and {Minniti}, Dante},
        title = "{Obscured star clusters in the inner Milky Way: How many massive young clusters are still awaiting detection}",
      journal = {\aap},
         year = 2024,
        month = dec,
       volume = {692},
          eid = {A194},
        pages = {A194},
          doi = {10.1051/0004-6361/202451078},
archivePrefix = {arXiv},
       eprint = {2411.02022},
 primaryClass = {astro-ph.GA},
       adsurl = {https://ui.adsabs.harvard.edu/abs/2024A&A...692A.194G}
}

@dataset{Hunt2023,
       author = {{Hunt}, E.~L. and {Reffert}, S.},
        title = "{VizieR Online Data Catalog: Improving the open cluster census. II. (Hunt+, 2023)}",
 howpublished = {VizieR On-line Data Catalog: J/A+A/673/A114. Originally published in: 2023A\&A...673A.114H},
         year = 2023,
        month = mar,
          eid = {J/A+A/673/A114},
       adsurl = {https://ui.adsabs.harvard.edu/abs/2023yCat..36730114H}
}

@ARTICLE{Hunt2024,
       author = {{Hunt}, Emily L. and {Reffert}, Sabine},
        title = "{Improving the open cluster census. III. Using cluster masses, radii, and dynamics to create a cleaned open cluster catalogue}",
      journal = {\aap},
         year = 2024,
        month = jun,
       volume = {686},
          eid = {A42},
        pages = {A42},
          doi = {10.1051/0004-6361/202348662},
archivePrefix = {arXiv},
       eprint = {2403.05143},
 primaryClass = {astro-ph.GA},
       adsurl = {https://ui.adsabs.harvard.edu/abs/2024A&A...686A..42H}
}

@ARTICLE{Ivanov2017,
       author = {{Ivanov}, Valentin D. and {Piatti}, Andr{\'e}s E. and {Beam{\'\i}n}, Juan-Carlos and {Minniti}, Dante and {Borissova}, Jordanka and {Kurtev}, Radostin and {Hempel}, Maren and {Saito}, Roberto K.},
        title = "{Candidate star clusters toward the inner Milky Way discovered on deep-stacked K$_{S}$-band images from the VVV Survey}",
      journal = {\aap},
         year = 2017,
        month = apr,
       volume = {600},
          eid = {A112},
        pages = {A112},
          doi = {10.1051/0004-6361/201630179},
archivePrefix = {arXiv},
       eprint = {1702.02394},
 primaryClass = {astro-ph.GA},
       adsurl = {https://ui.adsabs.harvard.edu/abs/2017A&A...600A.112I}
}

@ARTICLE{Jadhav2021,
       author = {{Jadhav}, Vikrant V. and {Roy}, Kaustubh and {Joshi}, Naman and {Subramaniam}, Annapurni},
        title = "{High Mass-Ratio Binary Population in Open Clusters: Segregation of Early Type Binaries and an Increasing Binary Fraction with Mass}",
      journal = {\aj},
         year = 2021,
        month = dec,
       volume = {162},
       number = {6},
          eid = {264},
        pages = {264},
          doi = {10.3847/1538-3881/ac2571},
archivePrefix = {arXiv},
       eprint = {2109.03782},
 primaryClass = {astro-ph.SR},
       adsurl = {https://ui.adsabs.harvard.edu/abs/2021AJ....162..264J}
}

@ARTICLE{Kastner1998,
       author = {{Kastner}, Joel H. and {Weintraub}, David A. and {Merrill}, K.~M. and {Gatley}, Ian},
        title = "{Direct Detection of the Mira at the Heart of OH 231.8+4.2}",
      journal = {\aj},
         year = 1998,
        month = sep,
       volume = {116},
       number = {3},
        pages = {1412-1418},
          doi = {10.1086/300520},
       adsurl = {https://ui.adsabs.harvard.edu/abs/1998AJ....116.1412K}
}

@ARTICLE{Kim2019,
       author = {{Kim}, Jaeheon and {Cho}, S.-H. and {Bujarrabal}, V. and {Imai}, H. and {Dodson}, R. and {Yoon}, D.-H. and {Zhang}, B.},
        title = "{Time variations of H$_{2}$O and SiO masers in the proto-Planetary Nebula OH 231.8+4.2}",
      journal = {\mnras},
         year = 2019,
        month = sep,
       volume = {488},
       number = {1},
        pages = {1427-1445},
          doi = {10.1093/mnras/stz1830},
archivePrefix = {arXiv},
       eprint = {1907.01825},
 primaryClass = {astro-ph.SR},
       adsurl = {https://ui.adsabs.harvard.edu/abs/2019MNRAS.488.1427K}
}

@article{King1962,
  author  = {King, I.},
  title   = {The structure of star clusters. I. an empirical density law},
  journal = {Astronomical Journal},
  volume  = {67},
  pages   = {471--485},
  year    = {1962},
  doi     = {10.1086/108756}
}

@ARTICLE{Kiss2008,
       author = {{Kiss}, L.~L. and {Szab{\'o}}, Gy. M. and {Balog}, Z. and {Parker}, Q.~A. and {Frew}, D.~J.},
        title = "{AAOmega radial velocities rule out current membership of the planetary nebula NGC 2438 in the open cluster M46}",
      journal = {\mnras},
         year = 2008,
        month = nov,
       volume = {391},
       number = {1},
        pages = {399-404},
          doi = {10.1111/j.1365-2966.2008.13899.x},
archivePrefix = {arXiv},
       eprint = {0809.0327},
 primaryClass = {astro-ph},
       adsurl = {https://ui.adsabs.harvard.edu/abs/2008MNRAS.391..399K}
}

@ARTICLE{Law2006,
       author = {{Law}, N.~M. and {Mackay}, C.~D. and {Baldwin}, J.~E.},
        title = "{Lucky imaging: high angular resolution imaging in the visible from the ground}",
      journal = {\aap},
         year = 2006,
        month = feb,
       volume = {446},
       number = {2},
        pages = {739-745},
          doi = {10.1051/0004-6361:20053695},
archivePrefix = {arXiv},
       eprint = {astro-ph/0507299},
 primaryClass = {astro-ph},
       adsurl = {https://ui.adsabs.harvard.edu/abs/2006A&A...446..739L}
}

@ARTICLE{Lodieu2009,
       author = {{Lodieu}, N. and {Zapatero Osorio}, M.~R. and {Mart{\'\i}n}, E.~L.},
        title = "{Lucky Imaging of M subdwarfs}",
      journal = {\aap},
         year = 2009,
        month = jun,
       volume = {499},
       number = {3},
        pages = {729-736},
          doi = {10.1051/0004-6361/200911708},
archivePrefix = {arXiv},
       eprint = {0903.4057},
 primaryClass = {astro-ph.GA},
       adsurl = {https://ui.adsabs.harvard.edu/abs/2009A&A...499..729L}
}

@ARTICLE{Majaess2007,
       author = {{Majaess}, Daniel J. and {Turner}, David G. and {Lane}, David J.},
        title = "{In Search of Possible Associations between Planetary Nebulae and Open Clusters}",
      journal = {\pasp},
         year = 2007,
        month = dec,
       volume = {119},
       number = {862},
        pages = {1349-1360},
          doi = {10.1086/524414},
archivePrefix = {arXiv},
       eprint = {0710.2900},
 primaryClass = {astro-ph},
       adsurl = {https://ui.adsabs.harvard.edu/abs/2007PASP..119.1349M}
}

@ARTICLE{Marigo2017,
       author = {{Marigo}, Paola and {Girardi}, L{\'e}o and {Bressan}, Alessandro and {Rosenfield}, Philip and {Aringer}, Bernhard and {Chen}, Yang and {Dussin}, Marco and {Nanni}, Ambra and {Pastorelli}, Giada and {Rodrigues}, Tha{\'\i}se S. and {Trabucchi}, Michele and {Bladh}, Sara and {Dalcanton}, Julianne and {Groenewegen}, Martin A.~T. and {Montalb{\'a}n}, Josefina and {Wood}, Peter R.},
        title = "{A New Generation of PARSEC-COLIBRI Stellar Isochrones Including the TP-AGB Phase}",
      journal = {\apj},
         year = 2017,
        month = jan,
       volume = {835},
       number = {1},
          eid = {77},
        pages = {77},
          doi = {10.3847/1538-4357/835/1/77},
archivePrefix = {arXiv},
       eprint = {1701.08510},
 primaryClass = {astro-ph.SR},
       adsurl = {https://ui.adsabs.harvard.edu/abs/2017ApJ...835...77M}
}

@ARTICLE{Majaess2012,
       author = {{Majaess}, D. and {Turner}, D. and {Moni Bidin}, C. and {Geisler}, D. and {Borissova}, J. and {Minniti}, D. and {Bonatto}, C. and {Gieren}, W. and {Carraro}, G. and {Kurtev}, R. and {Mauro}, F. and {Chen{\'e}}, A.-N. and {Forbes}, D. and {Lucas}, P. and {D{\'e}k{\'a}ny}, I. and {Saito}, R.~K. and {Soto}, M.},
        title = "{Strengthening the open cluster distance scale via VVV photometry}",
      journal = {\aap},
         year = 2012,
        month = jan,
       volume = {537},
          eid = {L4},
        pages = {L4},
          doi = {10.1051/0004-6361/201118614},
archivePrefix = {arXiv},
       eprint = {1112.3957},
 primaryClass = {astro-ph.GA},
       adsurl = {https://ui.adsabs.harvard.edu/abs/2012A&A...537L...4M}
}

@ARTICLE{Majaess2011,
       author = {{Majaess}, Daniel and {Turner}, David and {Moni Bidin}, Christian and {Mauro}, Francesco and {Geisler}, Douglas and {Gieren}, Wolfgang and {Minniti}, Dante and {Chen{\'e}}, Andr{\'e}-Nicolas and {Lucas}, Philip and {Borissova}, Jura and {Kurtev}, Radostn and {D{\'e}k{\'a}ny}, Istvan and {Saito}, Roberto K.},
        title = "{New Evidence Supporting Membership for TW Nor in Lyng{\r{a}} 6 and the Centaurus Spiral Arm}",
      journal = {\apjl},
         year = 2011,
        month = nov,
       volume = {741},
       number = {2},
          eid = {L27},
        pages = {L27},
          doi = {10.1088/2041-8205/741/2/L27},
archivePrefix = {arXiv},
       eprint = {1110.0830},
 primaryClass = {astro-ph.SR},
       adsurl = {https://ui.adsabs.harvard.edu/abs/2011ApJ...741L..27M}
}

@ARTICLE{Majaess2024,
       author = {{Majaess}, Daniel and {Turner}, David G. and {Minniti}, Dante and {Alonso-Garcia}, Javier and {Saito}, Roberto K.},
        title = "{The Valuable Long-period Cluster Cepheid KQ Scorpii and other Calibration Candidates}",
      journal = {\pasp},
         year = 2024,
        month = sep,
       volume = {136},
       number = {9},
          eid = {094202},
        pages = {094202},
          doi = {10.1088/1538-3873/ad7405},
archivePrefix = {arXiv},
       eprint = {2408.03371},
 primaryClass = {astro-ph.SR},
       adsurl = {https://ui.adsabs.harvard.edu/abs/2024PASP..136i4202M}
}

@ARTICLE{Majaess2025,
       author = {{Majaess}, Daniel and {Bonatto}, Charles J. and {Turner}, David G. and {Saito}, Roberto K. and {Minniti}, Dante and {Moni Bidin}, Christian and {Gonz{\'a}lez-D{\'\i}az}, Danilo and {Alonso-Garcia}, Javier and {Bono}, Giuseppe and {Braga}, Vittorio F. and {Navarro}, Maria G. and {Carraro}, Giovanni and {Gomez}, Matias},
        title = "{The Gaia Parallax Discrepancy for the Cluster Pismis 19 and Separating {\ensuremath{\delta}} Scutis from Cepheids}",
      journal = {\apj},
         year = 2025,
        month = apr,
       volume = {982},
       number = {2},
          eid = {165},
        pages = {165},
          doi = {10.3847/1538-4357/adb9e4},
archivePrefix = {arXiv},
       eprint = {2502.06930},
 primaryClass = {astro-ph.SR},
       adsurl = {https://ui.adsabs.harvard.edu/abs/2025ApJ...982..165M}
}

@ARTICLE{Martins2019,
       author = {{Martins}, F. and {Chen{\'e}}, A.-N. and {Bouret}, J.-C. and {Borissova}, J. and {Groh}, J. and {Ram{\'\i}rez Alegr{\'\i}a}, S. and {Minniti}, D.},
        title = "{Massive stars in the young cluster VVV CL074}",
      journal = {\aap},
         year = 2019,
        month = jul,
       volume = {627},
          eid = {A170},
        pages = {A170},
          doi = {10.1051/0004-6361/201935605},
archivePrefix = {arXiv},
       eprint = {1907.02357},
 primaryClass = {astro-ph.SR},
       adsurl = {https://ui.adsabs.harvard.edu/abs/2019A&A...627A.170M}
}

@ARTICLE{Matsuura2006,
       author = {{Matsuura}, M. and {Chesneau}, O. and {Zijlstra}, A.~A. and {Jaffe}, W. and {Waters}, L.~B.~F.~M. and {Yates}, J.~A. and {Lagadec}, E. and {Gledhill}, T. and {Etoka}, S. and {Richards}, A.~M.~S.},
        title = "{The Compact Circumstellar Material around OH 231.8+4.2}",
      journal = {\apjl},
         year = 2006,
        month = aug,
       volume = {646},
       number = {2},
        pages = {L123-L126},
          doi = {10.1086/507073},
archivePrefix = {arXiv},
       eprint = {astro-ph/0606576},
 primaryClass = {astro-ph},
       adsurl = {https://ui.adsabs.harvard.edu/abs/2006ApJ...646L.123M}
}

@ARTICLE{Mauro2012,
       author = {{Mauro}, Francesco and {Moni Bidin}, Christian and {Cohen}, Roger and {Geisler}, Doug and {Minniti}, Dante and {Catelan}, Marcio and {Chen{\'e}}, Andr{\'e}-Nicolas and {Villanova}, Sandro},
        title = "{Double Horizontal Branches in NGC 6440 and NGC 6569 Unveiled by the VVV Survey}",
      journal = {\apjl},
         year = 2012,
        month = dec,
       volume = {761},
       number = {2},
          eid = {L29},
        pages = {L29},
          doi = {10.1088/2041-8205/761/2/L29},
archivePrefix = {arXiv},
       eprint = {1211.3437},
 primaryClass = {astro-ph.SR},
       adsurl = {https://ui.adsabs.harvard.edu/abs/2012ApJ...761L..29M}
}

@ARTICLE{Meakin2003,
       author = {{Meakin}, C.~A. and {Bieging}, J.~H. and {Latter}, W.~B. and {Hora}, J.~L. and {Tielens}, A.~G.~G.~M.},
        title = "{Hubble Space Telescope/NICMOS Near-Infrared Imaging of the Proto-Planetary Nebula OH 231.8+4.2}",
      journal = {\apj},
         year = 2003,
        month = mar,
       volume = {585},
       number = {1},
        pages = {482-493},
          doi = {10.1086/345951},
       adsurl = {https://ui.adsabs.harvard.edu/abs/2003ApJ...585..482M}
}

@ARTICLE{Medina2018,
       author = {{Medina}, N. and {Borissova}, J. and {Bayo}, A. and {Kurtev}, R. and {Navarro Molina}, C. and {Kuhn}, M. and {Kumar}, N. and {Lucas}, P.~W. and {Catelan}, M. and {Minniti}, D. and {Smith}, L.~C.},
        title = "{An Automated Tool to Detect Variable Sources in the Vista Variables in the V{\'\i}a L{\'a}ctea Survey: The VVV Variables (V$^{4}$) Catalog of Tiles d001 and d002}",
      journal = {\apj},
         year = 2018,
        month = sep,
       volume = {864},
       number = {1},
          eid = {11},
        pages = {11},
          doi = {10.3847/1538-4357/aacc65},
archivePrefix = {arXiv},
       eprint = {1806.04061},
 primaryClass = {astro-ph.SR},
       adsurl = {https://ui.adsabs.harvard.edu/abs/2018ApJ...864...11M}
}

@ARTICLE{Medina2021,
       author = {{Medina}, N. and {Borissova}, J. and {Kurtev}, R. and {Alonso-Garc{\'\i}a}, J. and {Rom{\'a}n-Z{\'u}{\~n}iga}, Carlos G. and {Bayo}, A. and {Kounkel}, Marina and {Roman-Lopes}, Alexandre and {Lucas}, P.~W. and {Covey}, K.~R. and {F{\'o}rster}, Francisco and {Minniti}, Dante and {Adame}, Lucia and {Hern{\'a}ndez}, Jes{\'u}s},
        title = "{The G 305 Star-forming Region. II. Irregular Variable Stars}",
      journal = {\apj},
         year = 2021,
        month = jun,
       volume = {914},
       number = {1},
          eid = {28},
        pages = {28},
          doi = {10.3847/1538-4357/abf639},
archivePrefix = {arXiv},
       eprint = {2104.02200},
 primaryClass = {astro-ph.GA},
       adsurl = {https://ui.adsabs.harvard.edu/abs/2021ApJ...914...28M}
}

@ARTICLE{Minniti2010,
       author = {{Minniti}, D. and {Lucas}, P.~W. and {Emerson}, J.~P. and {Saito}, R.~K. and {Hempel}, M. and {Pietrukowicz}, P. and {Ahumada}, A.~V. and {Alonso}, M.~V. and {Alonso-Garcia}, J. and {Arias}, J.~I. and {Bandyopadhyay}, R.~M. and {Barb{\'a}}, R.~H. and {Barbuy}, B. and {Bedin}, L.~R. and {Bica}, E. and {Borissova}, J. and {Bronfman}, L. and {Carraro}, G. and {Catelan}, M. and {Clari{\'a}}, J.~J. and {Cross}, N. and {de Grijs}, R. and {D{\'e}k{\'a}ny}, I. and {Drew}, J.~E. and {Fari{\~n}a}, C. and {Feinstein}, C. and {Fern{\'a}ndez Laj{\'u}s}, E. and {Gamen}, R.~C. and {Geisler}, D. and {Gieren}, W. and {Goldman}, B. and {Gonzalez}, O.~A. and {Gunthardt}, G. and {Gurovich}, S. and {Hambly}, N.~C. and {Irwin}, M.~J. and {Ivanov}, V.~D. and {Jord{\'a}n}, A. and {Kerins}, E. and {Kinemuchi}, K. and {Kurtev}, R. and {L{\'o}pez-Corredoira}, M. and {Maccarone}, T. and {Masetti}, N. and {Merlo}, D. and {Messineo}, M. and {Mirabel}, I.~F. and {Monaco}, L. and {Morelli}, L. and {Padilla}, N. and {Palma}, T. and {Parisi}, M.~C. and {Pignata}, G. and {Rejkuba}, M. and {Roman-Lopes}, A. and {Sale}, S.~E. and {Schreiber}, M.~R. and {Schr{\"o}der}, A.~C. and {Smith}, M. and {Sodr{\'e}}, Jr., L. and {Soto}, M. and {Tamura}, M. and {Tappert}, C. and {Thompson}, M.~A. and {Toledo}, I. and {Zoccali}, M. and {Pietrzynski}, G.},
        title = "{VISTA Variables in the Via Lactea (VVV): The public ESO near-IR variability survey of the Milky Way}",
      journal = {\na},
         year = 2010,
        month = jul,
       volume = {15},
       number = {5},
        pages = {433-443},
          doi = {10.1016/j.newast.2009.12.002},
archivePrefix = {arXiv},
       eprint = {0912.1056},
 primaryClass = {astro-ph.GA},
       adsurl = {https://ui.adsabs.harvard.edu/abs/2010NewA...15..433M}
}

@ARTICLE{Molina2019,
       author = {{Molina}, J. and {Ibar}, Edo and {Villanueva}, V. and {Escala}, A. and {Cheng}, C. and {Baes}, M. and {Messias}, H. and {Yang}, C. and {Bauer}, F.~E. and {van der Werf}, P. and {Leiton}, R. and {Aravena}, M. and {Swinbank}, A.~M. and {Micha{\l}owski}, M.~J. and {Mu{\~n}oz-Arancibia}, A.~M. and {Orellana}, G. and {Hughes}, T.~M. and {Farrah}, D. and {De Zotti}, G. and {Lara-L{\'o}pez}, M.~A. and {Eales}, S. and {Dunne}, L.},
        title = "{VALES V: a kinematic analysis of the molecular gas content in H-ATLAS galaxies at z {\ensuremath{\sim}} 0.03-0.35 using ALMA}",
      journal = {\mnras},
         year = 2019,
        month = jan,
       volume = {482},
       number = {2},
        pages = {1499-1524},
          doi = {10.1093/mnras/sty2577},
archivePrefix = {arXiv},
       eprint = {1809.10752},
 primaryClass = {astro-ph.GA},
       adsurl = {https://ui.adsabs.harvard.edu/abs/2019MNRAS.482.1499M}
}

@ARTICLE{Minniti2017,
       author = {{Minniti}, Dante and {Palma}, Tali and {D{\'e}k{\'a}ny}, Istvan and {Hempel}, Maren and {Rejkuba}, Marina and {Pullen}, Joyce and {Alonso-Garc{\'\i}a}, Javier and {Barb{\'a}}, Rodolfo and {Barbuy}, Beatriz and {Bica}, Eduardo and {Bonatto}, Charles and {Borissova}, Jura and {Catelan}, Marcio and {Carballo-Bello}, Julio A. and {Chene}, Andre Nicolas and {Clari{\'a}}, Juan Jos{\'e} and {Cohen}, Roger E. and {Contreras Ramos}, Rodrigo and {Dias}, Bruno and {Emerson}, Jim and {Froebrich}, Dirk and {Buckner}, Anne S.~M. and {Geisler}, Douglas and {Gonzalez}, Oscar A. and {Gran}, Felipe and {Hajdu}, Gergely and {Irwin}, Mike and {Ivanov}, Valentin D. and {Kurtev}, Radostin and {Lucas}, Philip W. and {Majaess}, Daniel and {Mauro}, Francesco and {Moni-Bidin}, Christian and {Navarrete}, Camila and {Ram{\'\i}rez Alegr{\'\i}a}, Sebastian and {Saito}, Roberto K. and {Valenti}, Elena and {Zoccali}, Manuela},
        title = "{FSR 1716: A New Milky Way Globular Cluster Confirmed Using VVV RR Lyrae Stars}",
      journal = {\apjl},
         year = 2017,
        month = mar,
       volume = {838},
       number = {1},
          eid = {L14},
        pages = {L14},
          doi = {10.3847/2041-8213/838/1/L14},
archivePrefix = {arXiv},
       eprint = {1703.02033},
 primaryClass = {astro-ph.GA},
       adsurl = {https://ui.adsabs.harvard.edu/abs/2017ApJ...838L..14M}
}

@ARTICLE{Minniti2018,
       author = {{Minniti}, Dante and {Schlafly}, E.~F. and {Palma}, Tali and {Clari{\'a}}, Juan J. and {Hempel}, Maren and {Alonso-Garc{\'\i}a}, Javier and {Bica}, Eduardo and {Bonatto}, Charles and {Braga}, Vittorio F. and {Clementini}, Gisella and {Garofalo}, Alessia and {G{\'o}mez}, Mat{\'\i}as and {Ivanov}, Valentin D. and {Lucas}, Phillip W. and {Pullen}, Joyce and {Saito}, Roberto K. and {Smith}, Leigh C.},
        title = "{Confirmation of a New Metal-poor Globular Cluster in the Galactic Bulge}",
      journal = {\apj},
         year = 2018,
        month = oct,
       volume = {866},
       number = {1},
          eid = {12},
        pages = {12},
          doi = {10.3847/1538-4357/aadd06},
       adsurl = {https://ui.adsabs.harvard.edu/abs/2018ApJ...866...12M}
}

@ARTICLE{Morris1987,
       author = {{Morris}, Mark and {Guilloteau}, Stephane and {Lucas}, Robert and {Omont}, Alain},
        title = "{The Rich Molecular Spectrum and the Rapid Outflow of OH 231.8+4.2}",
      journal = {\apj},
         year = 1987,
        month = oct,
       volume = {321},
        pages = {888},
          doi = {10.1086/165681},
       adsurl = {https://ui.adsabs.harvard.edu/abs/1987ApJ...321..888M}
}

@ARTICLE{Navarro2016,
       author = {{Navarro Molina}, Claudio and {Borissova}, J. and {Catelan}, M. and {Alonso-Garc{\'\i}a}, J. and {Kerins}, E. and {Kurtev}, R. and {Lucas}, P.~W. and {Medina}, N. and {Minniti}, D. and {D{\'e}k{\'a}ny}, I.},
        title = "{Variable stars in the Quintuplet stellar cluster with the VVV survey$^{★}$}",
      journal = {\mnras},
         year = 2016,
        month = oct,
       volume = {462},
       number = {2},
        pages = {1180-1191},
          doi = {10.1093/mnras/stw1613},
archivePrefix = {arXiv},
       eprint = {1607.01795},
 primaryClass = {astro-ph.SR},
       adsurl = {https://ui.adsabs.harvard.edu/abs/2016MNRAS.462.1180N}
}

@ARTICLE{Ottl2014,
       author = {{{\"O}ttl}, S. and {Kimeswenger}, S. and {Zijlstra}, A.~A.},
        title = "{Ionization structure of multiple-shell planetary nebulae. I. NGC 2438}",
      journal = {\aap},
         year = 2014,
        month = may,
       volume = {565},
          eid = {A87},
        pages = {A87},
          doi = {10.1051/0004-6361/201323205},
archivePrefix = {arXiv},
       eprint = {1403.6715},
 primaryClass = {astro-ph.SR},
       adsurl = {https://ui.adsabs.harvard.edu/abs/2014A&A...565A..87O}
}

@ARTICLE{Palma2016,
       author = {{Palma}, T. and {Minniti}, D. and {D{\'e}k{\'a}ny}, I. and {Clari{\'a}}, J.~J. and {Alonso-Garc{\'\i}a}, J. and {Gramajo}, L.~V. and {Ram{\'\i}rez Alegr{\'\i}a}, S. and {Bonatto}, C.},
        title = "{New variable stars discovered in the fields of three Galactic open clusters using the VVV survey}",
      journal = {\na},
         year = 2016,
        month = nov,
       volume = {49},
        pages = {50-62},
          doi = {10.1016/j.newast.2016.05.008},
archivePrefix = {arXiv},
       eprint = {1606.05028},
 primaryClass = {astro-ph.SR},
       adsurl = {https://ui.adsabs.harvard.edu/abs/2016NewA...49...50P}
}

@ARTICLE{Palma2019,
       author = {{Palma}, Tali and {Minniti}, Dante and {Alonso-Garc{\'\i}a}, Javier and {Crestani}, Juliana and {Netzel}, Henryka and {Clari{\'a}}, Juan J. and {Saito}, Roberto K. and {Dias}, Bruno and {Fern{\'a}ndez-Trincado}, Jos{\'e} G. and {Kammers}, Roberto and {Geisler}, Douglas and {G{\'o}mez}, Mat{\'\i}as and {Hempel}, Maren and {Pullen}, Joyce},
        title = "{Analysis of the physical nature of 22 New VVV Survey Globular Cluster candidates in the Milky Way bulge}",
      journal = {\mnras},
         year = 2019,
        month = aug,
       volume = {487},
       number = {3},
        pages = {3140-3149},
          doi = {10.1093/mnras/stz1489},
archivePrefix = {arXiv},
       eprint = {1905.11835},
 primaryClass = {astro-ph.GA},
       adsurl = {https://ui.adsabs.harvard.edu/abs/2019MNRAS.487.3140P}
}

@ARTICLE{Pena2022,
       author = {{Pe{\~n}a Ram{\'\i}rez}, K. and {Smith}, L.~C. and {Ram{\'\i}rez Alegr{\'\i}a}, S. and {Chen{\'e}}, A.-N. and {Gonz{\'a}lez-Fern{\'a}ndez}, C. and {Lucas}, P.~W. and {Minniti}, D.},
        title = "{The VVV open cluster project - II. Near-infrared sequences of 37 open clusters on eight-dimensional parameter space}",
      journal = {\mnras},
         year = 2022,
        month = jul,
       volume = {513},
       number = {4},
        pages = {5799-5813},
          doi = {10.1093/mnras/stac1296},
archivePrefix = {arXiv},
       eprint = {2205.02735},
 primaryClass = {astro-ph.GA},
       adsurl = {https://ui.adsabs.harvard.edu/abs/2022MNRAS.513.5799P}
}

@ARTICLE{Ramirez2014,
       author = {{Ram{\'\i}rez Alegr{\'\i}a}, S. and {Borissova}, J. and {Chen{\'e}}, A.~N. and {O'Leary}, E. and {Amigo}, P. and {Minniti}, D. and {Saito}, R.~K. and {Geisler}, D. and {Kurtev}, R. and {Hempel}, M. and {Gromadzki}, M. and {Clarke}, J.~R.~A. and {Negueruela}, I. and {Marco}, A. and {Fierro}, C. and {Bonatto}, C. and {Catelan}, M.},
        title = "{Massive open star clusters using the VVV survey. III. A young massive cluster at the far edge of the Galactic bar}",
      journal = {\aap},
         year = 2014,
        month = apr,
       volume = {564},
          eid = {L9},
        pages = {L9},
          doi = {10.1051/0004-6361/201322619},
archivePrefix = {arXiv},
       eprint = {1403.3428},
 primaryClass = {astro-ph.GA},
       adsurl = {https://ui.adsabs.harvard.edu/abs/2014A&A...564L...9R}
}

@ARTICLE{Ray2022,
       author = {{Ray}, Amy E. and {Frinchaboy}, Peter M. and {Donor}, John and {Chojnowski}, S.~D. and {Melendez}, Matthew},
        title = "{The Open Cluster Chemical Abundances and Mapping Survey. V. Chemical Abundances of CTIO/Hydra Clusters Using The Cannon}",
      journal = {\aj},
         year = 2022,
        month = may,
       volume = {163},
       number = {5},
          eid = {195},
        pages = {195},
          doi = {10.3847/1538-3881/ac5835},
archivePrefix = {arXiv},
       eprint = {2202.05759},
 primaryClass = {astro-ph.GA},
       adsurl = {https://ui.adsabs.harvard.edu/abs/2022AJ....163..195R}
}

@ARTICLE{Reipurth1987,
       author = {{Reipurth}, Bo},
        title = "{Shocked bipolar outflow from the evolved star OH231.8 + 4.2}",
      journal = {\nat},
         year = 1987,
        month = feb,
       volume = {325},
       number = {6107},
        pages = {787-790},
          doi = {10.1038/325787a0},
       adsurl = {https://ui.adsabs.harvard.edu/abs/1987Natur.325..787R}
}

@ARTICLE{Sahai2005,
       author = {{Sahai}, R. and {Le Mignant}, D. and {S{\'a}nchez Contreras}, C. and {Campbell}, R.~D. and {Chaffee}, F.~H.},
        title = "{Sculpting a Pre-planetary Nebula with a Precessing Jet: IRAS 16342-3814}",
      journal = {\apjl},
         year = 2005,
        month = mar,
       volume = {622},
       number = {1},
        pages = {L53-L56},
          doi = {10.1086/429586},
       adsurl = {https://ui.adsabs.harvard.edu/abs/2005ApJ...622L..53S}
}

@ARTICLE{Saito2012,
       author = {{Saito}, R.~K. and {Hempel}, M. and {Minniti}, D. and {Lucas}, P.~W. and {Rejkuba}, M. and {Toledo}, I. and {Gonzalez}, O.~A. and {Alonso-Garc{\'\i}a}, J. and {Irwin}, M.~J. and {Gonzalez-Solares}, E. and {Hodgkin}, S.~T. and {Lewis}, J.~R. and {Cross}, N. and {Ivanov}, V.~D. and {Kerins}, E. and {Emerson}, J.~P. and {Soto}, M. and {Am{\^o}res}, E.~B. and {Gurovich}, S. and {D{\'e}k{\'a}ny}, I. and {Angeloni}, R. and {Beamin}, J.~C. and {Catelan}, M. and {Padilla}, N. and {Zoccali}, M. and {Pietrukowicz}, P. and {Moni Bidin}, C. and {Mauro}, F. and {Geisler}, D. and {Folkes}, S.~L. and {Sale}, S.~E. and {Borissova}, J. and {Kurtev}, R. and {Ahumada}, A.~V. and {Alonso}, M.~V. and {Adamson}, A. and {Arias}, J.~I. and {Bandyopadhyay}, R.~M. and {Barb{\'a}}, R.~H. and {Barbuy}, B. and {Baume}, G.~L. and {Bedin}, L.~R. and {Bellini}, A. and {Benjamin}, R. and {Bica}, E. and {Bonatto}, C. and {Bronfman}, L. and {Carraro}, G. and {Chen{\`e}}, A.~N. and {Clari{\'a}}, J.~J. and {Clarke}, J.~R.~A. and {Contreras}, C. and {Corvill{\'o}n}, A. and {de Grijs}, R. and {Dias}, B. and {Drew}, J.~E. and {Fari{\~n}a}, C. and {Feinstein}, C. and {Fern{\'a}ndez-Laj{\'u}s}, E. and {Gamen}, R.~C. and {Gieren}, W. and {Goldman}, B. and {Gonz{\'a}lez-Fern{\'a}ndez}, C. and {Grand}, R.~J.~J. and {Gunthardt}, G. and {Hambly}, N.~C. and {Hanson}, M.~M. and {He{\l}miniak}, K.~G. and {Hoare}, M.~G. and {Huckvale}, L. and {Jord{\'a}n}, A. and {Kinemuchi}, K. and {Longmore}, A. and {L{\'o}pez-Corredoira}, M. and {Maccarone}, T. and {Majaess}, D. and {Mart{\'\i}n}, E.~L. and {Masetti}, N. and {Mennickent}, R.~E. and {Mirabel}, I.~F. and {Monaco}, L. and {Morelli}, L. and {Motta}, V. and {Palma}, T. and {Parisi}, M.~C. and {Parker}, Q. and {Pe{\~n}aloza}, F. and {Pietrzy{\'n}ski}, G. and {Pignata}, G. and {Popescu}, B. and {Read}, M.~A. and {Rojas}, A. and {Roman-Lopes}, A. and {Ruiz}, M.~T. and {Saviane}, I. and {Schreiber}, M.~R. and {Schr{\"o}der}, A.~C. and {Sharma}, S. and {Smith}, M.~D. and {Sodr{\'e}}, L. and {Stead}, J. and {Stephens}, A.~W. and {Tamura}, M. and {Tappert}, C. and {Thompson}, M.~A. and {Valenti}, E. and {Vanzi}, L. and {Walton}, N.~A. and {Weidmann}, W. and {Zijlstra}, A.},
        title = "{VVV DR1: The first data release of the Milky Way bulge and southern plane from the near-infrared ESO public survey VISTA variables in the V{\'\i}a L{\'a}ctea}",
      journal = {\aap},
         year = 2012,
        month = jan,
       volume = {537},
          eid = {A107},
        pages = {A107},
          doi = {10.1051/0004-6361/201118407},
archivePrefix = {arXiv},
       eprint = {1111.5511},
 primaryClass = {astro-ph.GA},
       adsurl = {https://ui.adsabs.harvard.edu/abs/2012A&A...537A.107S}
}

\clearpage

\begin{appendix}
\section{Comparison with the deep stacks}\label{appendix}
\begin{figure}[h]
  \begin{center}
    \includegraphics[width=82mm]{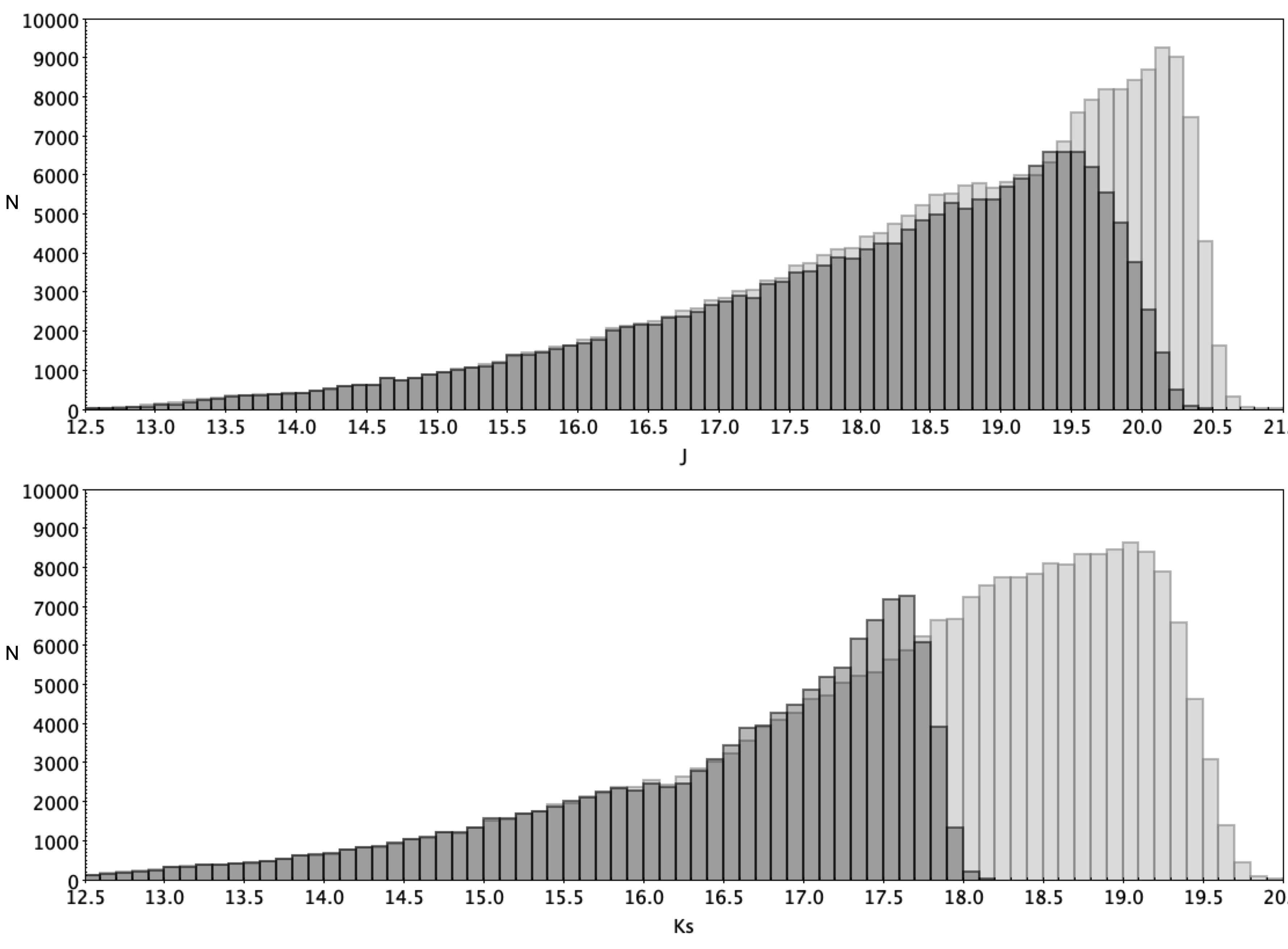}
    \caption{Near-IR  magnitude distributions of detected sources in single vs. deep stacked images. Bottom: $K_s$-band magnitude distribution of point sources detected using the DoPhot PSF photometry of a single chip for VVVX tile e0851. The PSF photometry for new deep stacked images from the VVVX Deep Stacks (shown in light gray) reach about 1.6 mag deeper in the $K_s$-band than the previous catalogs available from single images (in black). This results in the detection of 121420 more point sources, a significant improvement. Top: The PSF photometry for new deep stacked images from the VVVX Deep Stacks (shown in light gray) reach about 0.5 mag deeper in the $J$-band than the previously available data (in black).  }
    \label{FigureA1}
  \end{center}
\end{figure}

In this section, we briefly compare the VVVX near-IR PSF photometry of single epochs compared with those of the deep stacks, in order to illustrate the progress that has been made. As an example of the improved gain in depth and completeness, we show in Fig. \ref{FigureA1} the PSF magnitude distributions of detected sources in single images compared to the deep stacks of VVVX tile e0851 in the $J$ and $K_s$ band filters, respectively.

\begin{figure}[h]
  \begin{center}
    \includegraphics[width=82mm]{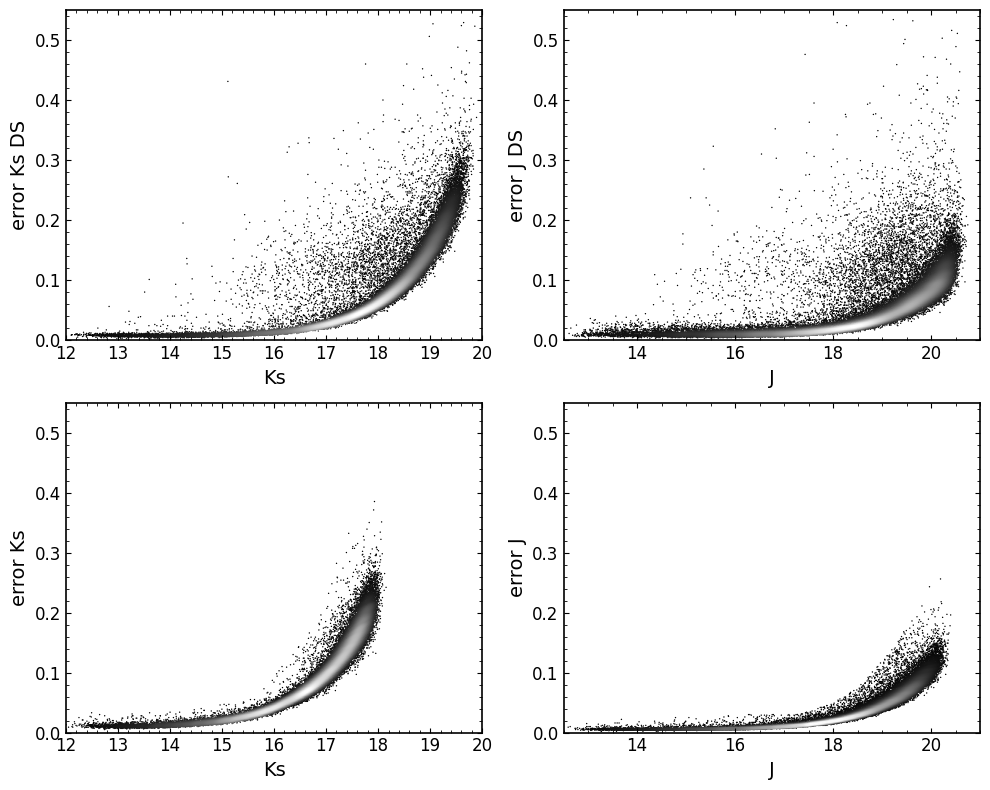}
    \caption{Top panels: Errors for the near-IR PSF photometry as a function of magnitude for new images from the VVVX Deep Stacks. Bottom panels: Errors for the previous near-IR PSF photometry as function of magnitude for single epoch images from \citet{Alonso2018}. }
    \label{FigureA2}
  \end{center}
\end{figure}

Fig. \ref{FigureA2} shows the errors for the near-IR PSF photometry as a function of magnitude for new images from the VVVX deep stacks (top panels) compared to the single epochs (bottom panels). 

\begin{figure}[h]
  \begin{center}
    \includegraphics[width=82mm]{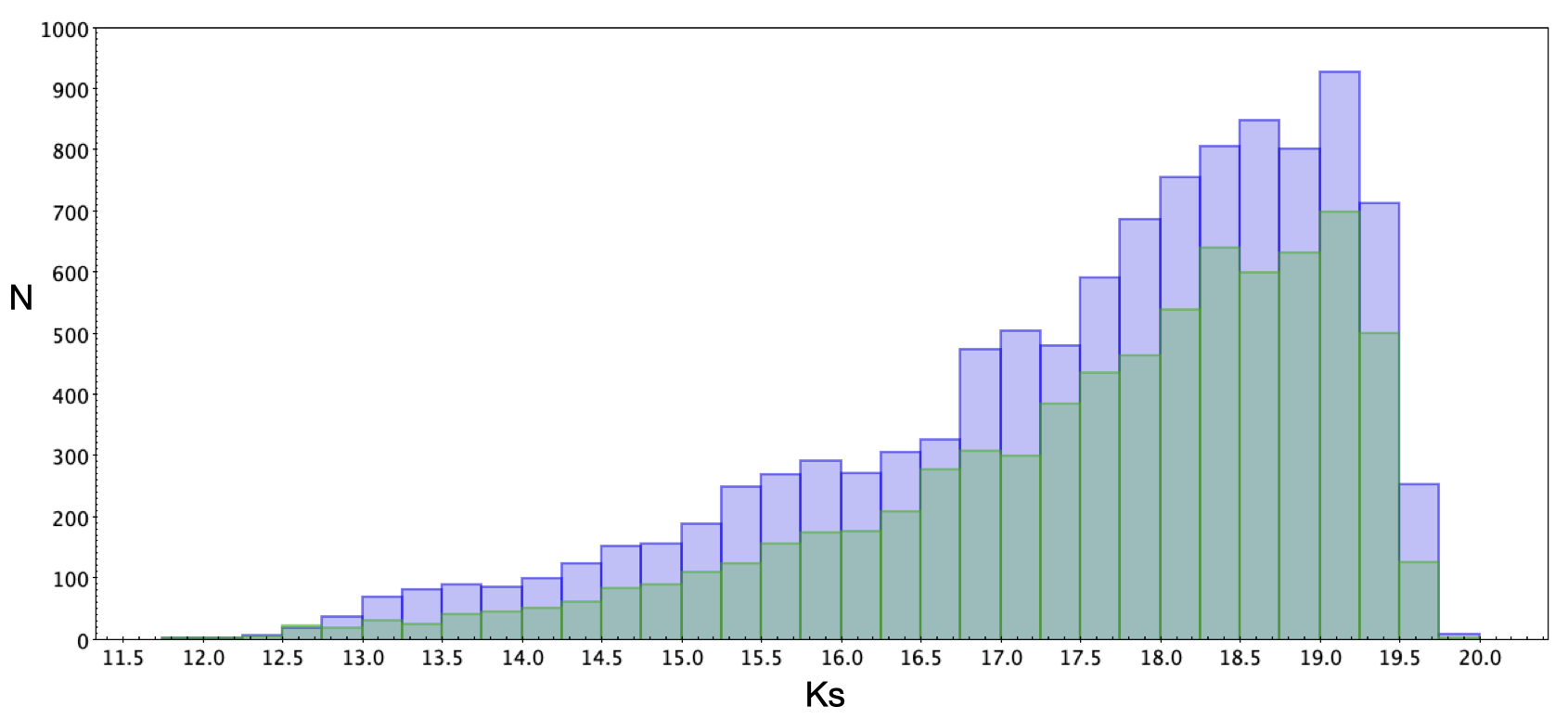}
    \caption{ Near-IR  magnitude distribution of detected sources in the deep stacked images for the region of the cluster NGC\,2437 compared with a selected background field. }
    \label{FigureA3}
  \end{center}
\end{figure}

Finally, for the specific case of the open cluster benchmark NGC\,2437, Fig. \ref{FigureA3} shows the $K_s$ band magnitude distribution of detected sources in the deep stacked images for the region of the cluster NGC\,2437 compared with a selected background field located about one degree away.
This figure shows that there is an excess of cluster stars even at the faintest magnitudes $K_s = 19.5$ mag, equivalent to absolute magnitude $M_{K_s} = 8.6$, near the BD boundary.

\end{appendix}

\end{document}